# A Sub-Neptune Exoplanet with a Low-Metallicity Methane-Depleted Atmosphere and Mie-Scattering Clouds

BJÖRN BENNEKE[1], HEATHER A. KNUTSON[2], JOSHUA LOTHRINGER[3], IAN J.M. CROSSFIELD[4], JULIANNE I. MOSES[5], CAROLINE MORLEY[6], LAURA KREIDBERG[7], BENJAMIN J. FULTON[2,17], DIANA DRAGOMIR[4,18], ANDREW W. HOWARD[8], IAN WONG[9], JEAN-MICHEL DÉSERT[10], PETER R. MCCULLOUGH[11],ELIZA M.-R. KEMPTON[12,13,], JONATHAN FORTNEY[14], RONALD GILLILAND[15], DRAKE DEMING[12], JOSHUA KAMMER[16]

[1] Department of Physics and Institute for Research on Exoplanets, Université de Montréal, Montréal, QC, Canada
[2] Division of Geological and Planetary Sciences, California Institute of Technology, Pasadena, CA 91125, USA
[3] Lunar & Planetary Laboratory, University of Arizona, 1629 E. University Boulevard., Tucson, AZ, USA
[4] Department of Physics and Kavli Institute of Astronomy, Massachusetts Institute of Technology, 77 Massachusetts Ave, Cambridge, MA, 02139, USA
[5] Space Science Institute, 4750 Walnut Street, Suite 205, Boulder, CO 80301, USA
[6] Department of Astronomy, University of Texas, Austin, TX 78712, USA
[7] Department of Astronomy, Harvard University, 60 Garden Street, Cambridge, MA 02138, USA
[8] Department of Astronomy, California Institute of Technology, Pasadena, CA 91125, USA
[9] Department of Earth, Atmospheric, and Planetary Sciences, Massachusetts Institute of Technology, 77 Massachusetts Ave, Cambridge, MA, 02139, USA
[10] Anton Pannekoek Institute for Astronomy, University of Amsterdam, 1090 GE Amsterdam, The Netherlands
[11] Department of Physics and Astronomy, Johns Hopkins University, Baltimore, MD 21218, USA
[12] Department of Astronomy, University of Maryland, College Park, MD 20742, USA
[13] Department of Physics, Grinnell College, 1116 8th Avenue, Grinnell, IA 50112, USA
[14] Department of Astronomy, University of California, Santa Cruz, CA 95064, USA
[15] Space Telescope Science Institute, 3700 San Martin Dr., Baltimore, MD 21218, USA
[16] Southwest Research Institute, San Antonio TX, USA
[17] IPAC-NASA Exoplanet Science Institute Pasadena, CA 91125, USA
[18] NASA Hubble Fellow

**With no analogues in the Solar System, the discovery of thousands of exoplanets with masses and radii intermediate between Earth and Neptune was one of the big surprises of exoplanet science. These super-Earths and sub-Neptunes likely represent the most common outcome of planet formation[1,2]. Mass and radius measurements indicate a diversity in bulk composition much wider than for gas giants[3]; however, direct spectroscopic detections of molecular absorption and constraints on the gas mixing ratios have largely remained limited to planets more massive than Neptune[4–6]. Here, we analyze a combined Hubble/Spitzer Space Telescope dataset of 12 transits and 20 eclipses of the sub-Neptune GJ 3470 b, whose mass of 12.6 $M_\oplus$ places it near the half-way point between previously studied exo-Neptunes (22-23 $M_\oplus$)[5–7] and exoplanets known to have rocky densities (7 $M_\oplus$)[8]. Obtained over many years, our data set provides a robust detection of water absorption (>5σ) and a thermal emission detection from the lowest irradiated planet to date. We reveal a low-metallicity, hydrogen-dominated atmosphere similar to a gas giant, but strongly depleted in methane gas. The low, near-solar metallicity (O/H=0.2-18) sets important constraints on the potential planet formation processes at low masses as well as the subsequent accretion of solids. The low methane abundance indicates that methane is destroyed much more efficiently than previously predicted, suggesting that the $CH_4$/CO transition curve has to be revisited for close-in planets. Finally, we also find a sharp drop in the cloud opacity at 2-3 μm characteristic of Mie scattering, which enables narrow constraints on the cloud particle size and makes GJ 3470b a keystone target for mid-IR characterization with JWST.**

We observed the transiting sub-Neptune mass exoplanet GJ 3470b with the Hubble Space Telescope (HST) as part of a spectral survey of atmospheres of low mass exoplanets (GO 13665). With an orbital period of 3.3 days and a mass of 12.6 $M_\oplus$, GJ 3470b is a typical member of the intriguingly abundant class of close-in sub-Neptunes (Figure 1). GJ 3470b's low surface gravity combined with the proximity and small size of its host star make it an outstanding candidate for detailed atmospheric characterization, especially in the sub-Neptune mass regime for which robust molecular detections have remained elusive to date[4,9,10]. We obtained time-series spectroscopy of six transits using HST, including three from 1.1 μm to 1.7 μm using Wide Field Camera 3 (WFC3) and three at optical wavelengths (0.55–1.0 μm) using the Space Telescope Imaging Spectrograph (STIS). We complement these HST transit observations with a total six Spitzer/IRAC transit observations as well as a total of 20 secondary eclipse observations at 3.6 and 4.5 μm (Figure 2 and 3).

We jointly analyze all HST and Spitzer transit data to obtain a consistent visible-to-IR transmission spectrum covering a wavelength range between 0.55 and 5.0 μm (Figure 2). The details of this analysis are described in the Methods Section. For each instrument, we verify that the measurements are consistent over multiple epochs by demonstrating repeatability among the three transits in each channel, setting tight upper limits on the effect of star spots on the overall spectrum (Supplementary Figure 4). The transit depth precision with WFC3 is substantially higher than the one obtained with STIS and Spitzer due to GJ 3470's higher photon flux in the WFC3 bandpass and the substantially higher throughput of WFC3 (30-40%) compared to STIS (8-12%). The achromaticity of the dominant WFC3 systematics further improves the transit depth precision in the WFC3 spectroscopic channels relative to STIS. Our three transit observations using optimized WFC3 spatial scans across the full sub-array enable us to collect 30 times more photons and achieve 5 times higher precision than the previously published stare-mode transit observation of GJ 3470b with this same instrument[9].

Our transmission spectrum for GJ 3470b reveals an attenuated but statistically significant water absorption feature at 1.4 μm in the WFC3 data, protruding over an otherwise cloud opacity-dominated visible to near-IR transmission spectrum (Figure 2). The water absorption is detected in multiple neighboring spectroscopic channels covering the 1.4 μm water band. Quantitively, retrieval models that include molecular absorption by water are favored by the Bayesian evidence[11] at 124,770:1 (5.2σ) and result in significantly better best fits than models without water (see Methods). Comparisons to models show that the data are best matched by a low-metallicity, hydrogen-dominated atmosphere (O/H = 0.2–18 x solar) with water vapor absorbing above high-altitude clouds that become optically thick below the 1 mbar level at 1.5μm (Figure 4). High metallicity atmospheres with high mean molecular mass are ruled out by the data because the associated smaller scale height would not allow for the observed transit depth variations. This finding is independent of the detailed atmospheric models because the observed transit depth variations would require the cut-off altitude of the grazing star light to vary over greater than 20 atmospheric scale heights across the near-infrared, which is not realistic. The Spitzer eclipse observations add to the water constraints because substantially increased water opacity would not allow for the observed contrast in thermal emission at 3.6 and 4.5 μm.

Intriguingly, the high-altitude clouds on GJ 3470b are not well represented by Rayleigh hazes as previously reported[12,13] or a simple gray cloud deck. Instead, our measurements provide direct observational evidence for the characteristic wavelength dependent extinction of finite-sized Mie scattering aerosol particles (Figure 2). These aerosol clouds become increasingly transparent at around 2–3 μm enabling us to constrain their effective particle size to 0.60±0.06 μm in the uppermost layers of the clouds (Figure 4). This particle size estimate provides a rare direct observational constraint that can guide the further development and verification of physics-driven cloud and haze models for exoplanets. Here, we account for the non-gray cloud opacities in our retrieval analysis by modeling the finite-sized particles using Mie scattering theory[11,14] and describing the effective particle size, the upper cloud deck pressure, and the cloud scale height as free parameters (see Methods). All posterior constraints on the atmospheric gases provided in this work account for the uncertainties introduced by the clouds as well as the parameterized "free" temperature structure (see Supplementary Figures 6 and 7).

GJ 3470b's transmission spectrum also shows a striking absence of methane absorption. For the relatively cool, low-metallicity atmosphere of GJ 3470b, atmosphere models with solar carbon-to-oxygen ratio would have predicted methane to be the dominant carbon-bearing molecule. However, methane absorption at 1.6μm in the WFC3 bandpass and at 3.3μm in Spitzer/IRAC channel 1 is not observed indicating a strong depleting of methane (Figure 2). The low methane abundance is independently supported by the twenty secondary eclipse observations, which can be used to constrain the shape of GJ 3470b's thermal emission spectrum in the 3.6 and 4.5 μm Spitzer bands (Figure 3). We detect strong thermal flux emerging at 3.6 μm at 4.7σ significance ($F_P/F_* = 115^{+27}_{-26}$ p.p.m) and a tight upper limit at 4.5 μm ($F_P/F_* = 3 \pm 22$ p.p.m). This is in contrast to the prediction for the fiducial solar abundance model, but in agreement with the low methane abundance inferred from the transmission spectrum (Figure 2). Quantitatively, our retrieval analysis shows that the methane abundance is below $1.3 \times 10^{-5}$ at greater than 99.7% confidence, substantially below the value of $4.6 \times 10^{-4}$ expected for a solar abundance atmosphere in chemical equilibrium (Figure 4). The best fitting models show a striking methane depletion by three orders of magnitude compared to equilibrium.

We assess the origin of the methane depletion through state-of-the-art photochemical modeling and thermal modeling of GJ 3470b's atmosphere (see Methods). Consistent with Refs [15,16] we find that the methane abundance in the layers probed by the observations should not be reduced substantially by photochemistry in layers probed by our observations (Figure 4, Supplementary Figure 8). Possible explanations for the unexpected lack of methane could be substantial interior heating, photochemical depletion due to catalytic destruction of $CH_4$ in deeper atmospheric regions, or a low C/O ratio as a result of the planet formation process. The interior heating scenario would require interior temperatures ($T_{int}$) above 300 K to push the otherwise relatively cold mid-atmosphere of GJ 3470b into the CO dominated regime[17,18]. Evolution modeling of GJ 3470b indicates that internal heat from formation should have been radiated away within a few Myr, well below the estimated age of the system[19]; however, tidal heating due to forced eccentricity from another unseen planet in the system, similar to the situation with Jupiter's moon Io could be a possible explanation. The residual non-zero eccentricity of GJ 3470b as independently confirmed by our eclipse observations and radial velocity measurements support this hypothesis. Alternatively, GJ 3470b's surprising lack of methane could potentially be the results of photochemical depletion due to catalytic destruction of $CH_4$ in deeper atmospheric regions where photolysis of $NH_3$ and $H_2S$ release large amounts of atomic hydrogen. The fact that ammonia is also depleted in comparison to expectations based on our chemical-kinetics modeling (Figure 4) is consistent with this catalytic-destruction possibility. Ammonia is an important quenched disequilibrium product (Supplementary Figure 8), and we would have expected to see $NH_3$ absorption at 1.5 μm (Figure 2). In either scenario, the carbon freed from methane would most likely be locked up in CO, which can be seen in the transmission spectrum at 4.5μm (Figure 2) and as a suppression of thermal flux within the 4.5μm Spitzer bandpass (Figure 3). HCN is also one potential major sink of the carbon in the coupled $CH_4$-$NH_3$ photochemistry if the elemental N/C ratio is larger than solar[20,21]; however, we also obtain an upper bound on the HCN abundances (Supplementary Figure 7). In either case, all $CH_4$ destruction scenarios would also lead to the production of $CH_3$ and other radicals, some fraction of which can react with other atmospheric carbon and nitrogen species to form increasingly heavy hydrocarbons and nitriles, eventually ending up in refractory soot-like haze particles. These photochemically produced particles could provide an explanation for the observed cloud opacity at shortward of 3 μm, although recent experimental work has also shown that similar photochemical hazes can be formed even in the absence of methane[22,23]. In either case, the 30-90 nm particles found in experiments[2] would likely need to coagulate to form larger aggregates to explain the inferred wavelength dependence of the cloud opacity on GJ 3470b.

Overall, our spectra show directly through atmospheric observations that close-in sub-Neptunes can have near-solar metallicity atmospheres likely formed by the direct accretion of primordial gas from the protoplanetary disk onto a rock/iron or ice-dominated core as suggested by recent planet formation models[19,24]. The near-solar metallicity is particularly intriguing because the steep increase in planet occurrence rate from >20 $M_\oplus$ towards 10 $M_\oplus$[25] suggests that sub-Neptunes could have a more efficient planet formation process intrinsically different from planets more massive than Neptune[4–7,10,17,26]. Sub-Neptune formation beyond the ice line and subsequent migration could have led to much higher atmospheric metallicities, or even water worlds[27], which we do not find for GJ 3470b. Instead, our measurement of a near-solar water abundance favors formation scenarios in which the core accreted a primordial gas envelope whose metal content was subsequently not notably enriched by planetesimal accretion[28], the erosion of the icy/rocky core, or late collisions[29]. GJ 3470b's gas envelope is also expected to have undergone substantial mass loss[30], but this mass loss was likely in the hydrodynamic regime as we see no evidence for preferential loss of lighter elements. Finally, the unexpected methane depletion further indicates that our understanding of the chemical and thermal processes on these low-mass planets remains incomplete. Opportunely, our detection of sub-μm Mie scattering particles indicates that the clouds or hazes in the atmosphere of GJ 3470b become increasingly transparent beyond 3 μm, making GJ 3470b an excellent target for future JWST mid-IR observations—both as an archetype for the intriguing population of sub-Neptunes and as a laboratory for atmospheric chemistry and cloud formation in warm atmospheres.

**Acknowledgement.** This work is based on observations with the NASA/ESA HST, obtained at the Space Telescope Science Institute (STScI) operated by AURA, Inc. We received support for the analyze by NASA through grants under the HST-GO-13665 program (PI Benneke). This work is also based in part on observations made with the Spitzer Space Telescope, which is operated by the Jet Propulsion Laboratory, California Institute of Technology under a contract with NASA (PIs Knutson and Désert). B.B. further acknowledges financial supported by the Natural Sciences and Engineering Research Council (NSERC) of Canada and the Fond de Recherche Québécois—Nature et Technologie (FRQNT; Québec). J.M. acknowledges support from NASA grant NNX16AC64G, the Amsterdam Academic Alliance (AAA) Program, and European Research Council (ERC) under the programme Exo-Atmos (grant agreement no. 679633). D. D. acknowledges support provided by NASA through Hubble Fellowship grant HST-HF2-51372.001-A awarded by the Space Telescope Science Institute, which is operated by the Association of Universities for Research in Astronomy, Inc., for NASA, under contract NAS5-26555.

**Author contributions.** B.B. led the data analysis of the HST and Spitzer transit data, with contributions from J.L., I.W., and H.K. L.K. and J.M.D. performed independent analyses of the Spitzer transits and found consistent results. H.K. led the data analysis of the Spitzer secondary eclipse observations. J.M. provided the chemical kinetics atmosphere models. B.B. and C.M. provided the self-consistent atmospheric models and the atmospheric retrieval analysis. B.B. wrote the manuscript, with contributions from B.F., H.K. and J.M. All authors discussed the results and commented on the draft.

**Competing interests.** The authors declare no competing financial interests.

**Author information.** Correspondence and requests for materials should be addressed to B.B. (bbenneke@astro.umontreal.ca).

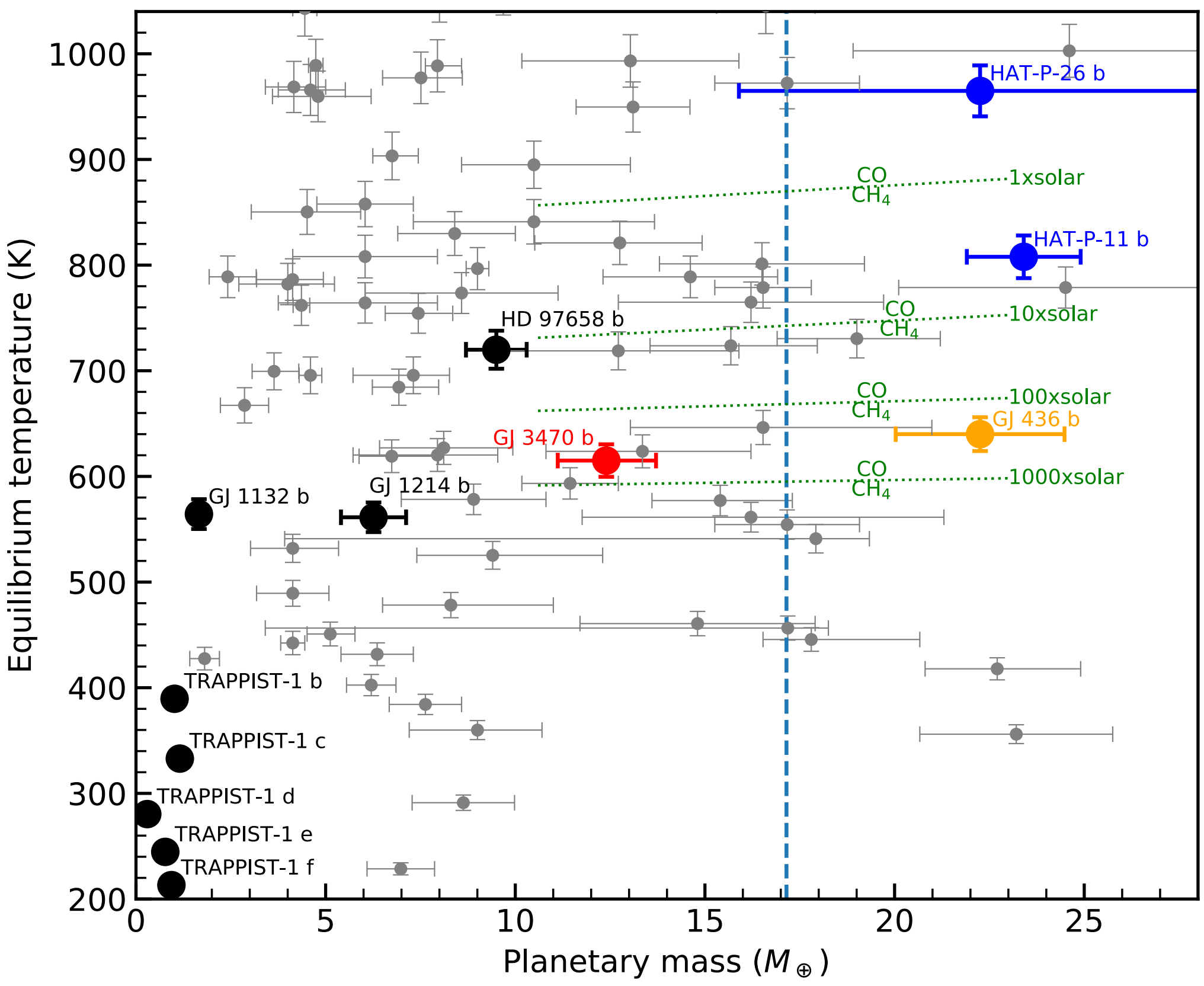

**Figure 1: Planet mass versus equilibrium temperature for known low-mass planets with signal-to-noise ratio >3 mass measurements.** Planets with published space-based spectroscopic observations for atmospheric characterization are indicated by large circles (black and coloured) and all other planets by small grey circles. Equilibrium temperatures are calculated for a Bond albedo of $A_B$ = 0.1. Vertical and horizontal bars indicate the 1σ uncertainties. Among the spectroscopically studied planets, the planets with non-detections of the atmosphere are shown in black, planets with detection of only transit features in blue, planets with detected thermal emission deviant from black-body radiation in yellow, and GJ 3470 b with spectral features detected both in transit and eclipse measurements in red. The green dotted curves show the equilibrium temperatures for which we expect $CH_4$ (below the curve) and CO (above the curve) to be the dominant carbon-bearing species in the photosphere based on self- consistent modelling in chemical equilibrium. The dashed vertical line indicates the mass of Neptune for reference.

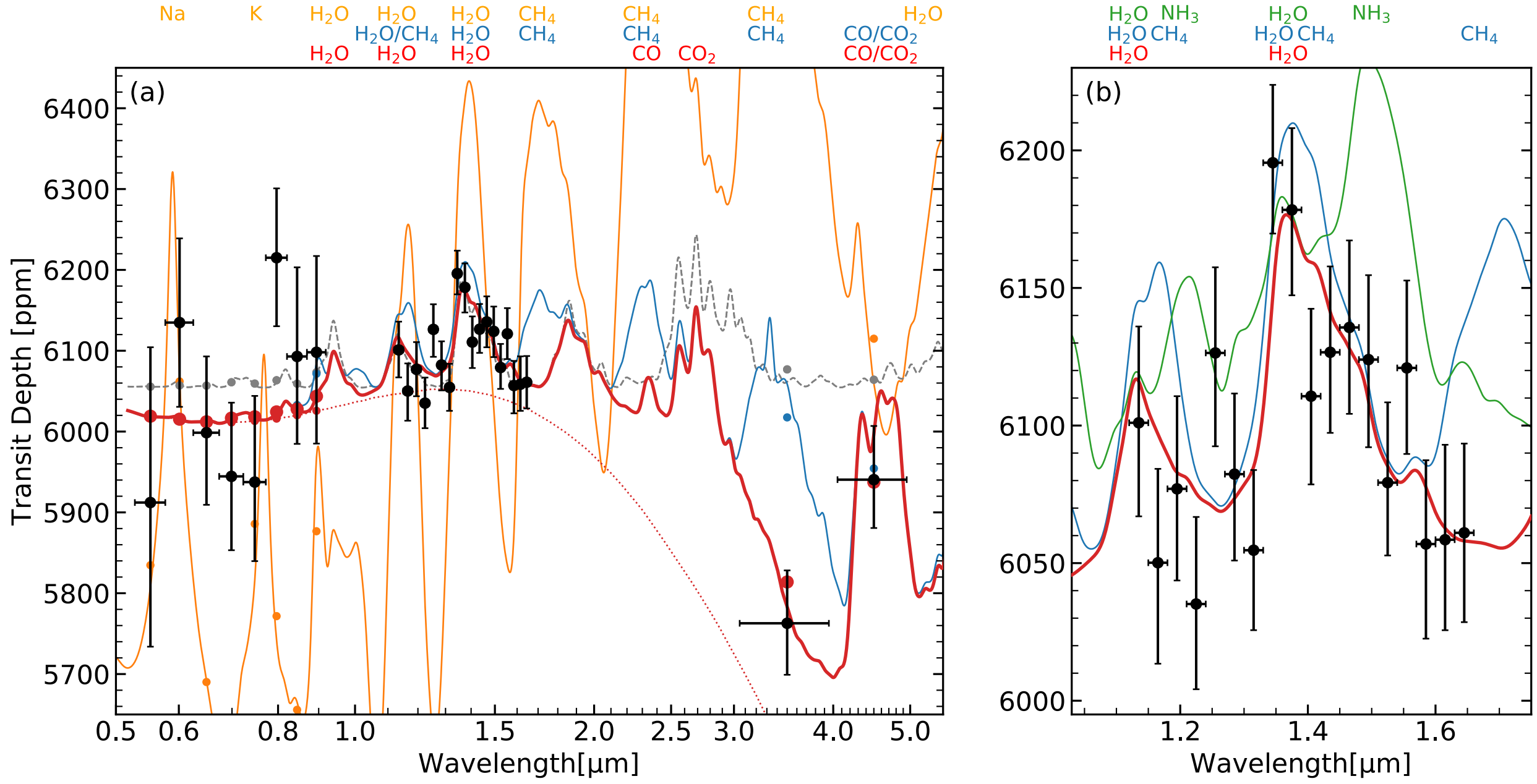

**Figure 2: Transmission spectrum of GJ 3470b.** Black data points show transit depth measurements from the HST/STIS, HST/WFC3, and Spitzer/IRAC observations analyzed in this study. Vertical and horizontal black bars indicate the 1σ transit depth uncertainties and the wavelength ranges of the measurements, respectively. The best fitting model with near-solar water abundance, Mie scattering clouds, and strong methane depletion is shown by the red curve, with circles indicating the bandpass integrated model. Water absorption results in increased transit depth at 1.4 μm (zoom in panel b). Finite-sized Mie-scattering particles (~0.6 μm) result in a characteristic drop off in cloud opacity beyond 2 μm (red dotted curved). Adding 100 ppm methane to the best fit model results in significant disagreement to the data at 1.6 and 3.6 μm (blue curve). Similarly, adding 100 ppm ammonia results in disagreement at 1.5 μm (green curve in panel b). A cloud-free solar metallicity model (orange curve) and the best-fitting gray cloud model (gray dashed curve) are shown in panel (a) for reference. Both provide a poor fit to the data. The dominant molecular absorbers for each model are labeled at the top with colors matching the color of the spectra. A distribution of models from the joint retrieval modeling of transit and eclipse data is shown in Supplementary Figure 6. Previous measurements[9,12,31–34] are in statistical agreement with our data, but have significantly larger transit depth uncertainties and are omitted here for clarity.

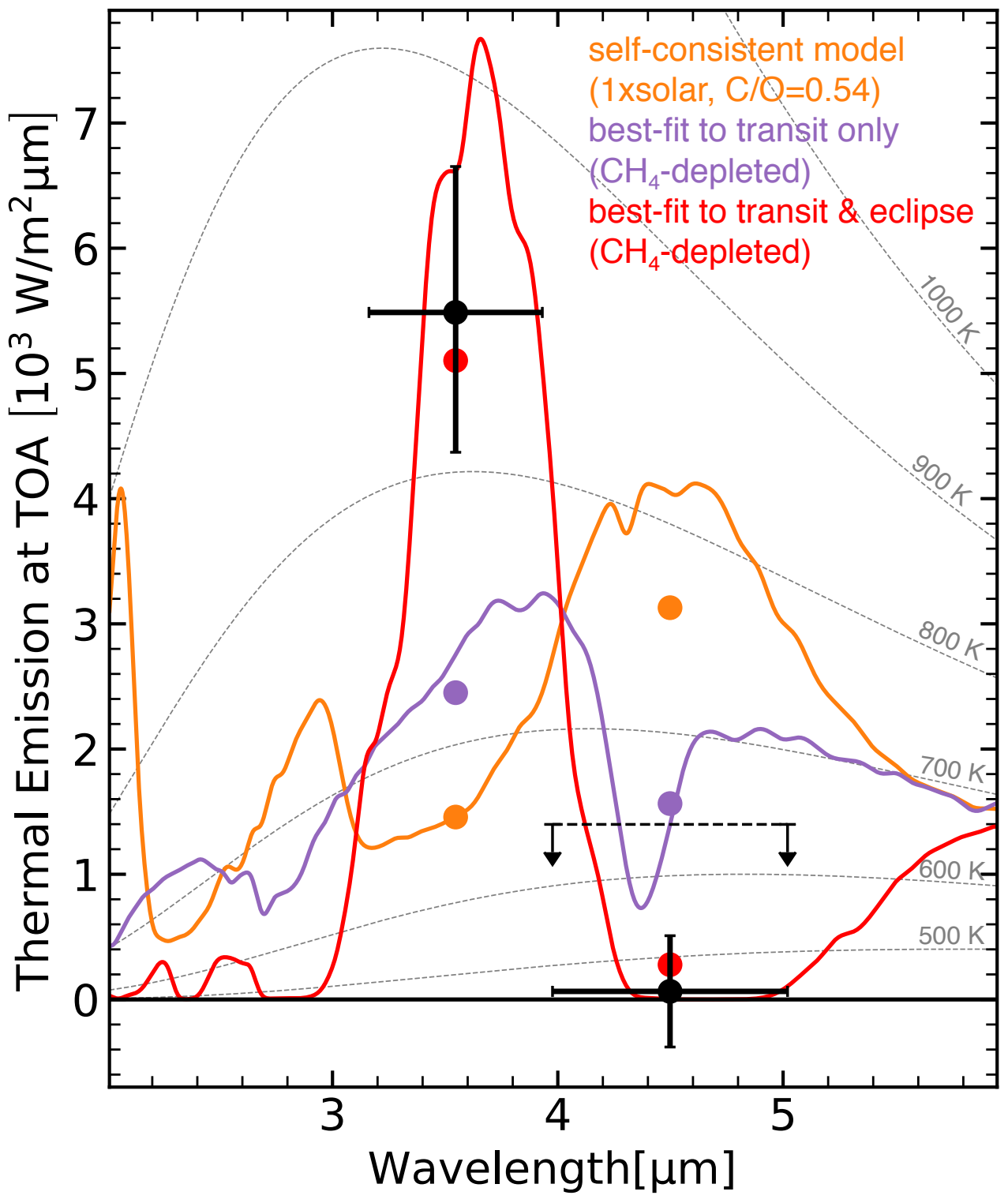

**Figure 3: Thermal emission spectrum of GJ 3470 b.** Spitzer/IRAC measurements (black) are compared to simulated model atmospheres (colored solid curves) and black body curves (grey dotted curves). Vertical and horizontal black bars indicate the 1σ transit depth uncertainties and the wavelength ranges of the measurements, respectively. The horizontal black dashed line represents the 3σ upper limit for the 4.5 μm measurement. Colored circles indicate the bandpass integrated models. Consistent with the transit data, the fiducial methane-rich, chemical equilibrium model with solar metallicity (orange) results in a poor fit to the data. Instead, the strong flux reversal between 3.6 μm and 4.5 μm Spitzer/IRAC bandpasses strongly favors methane-depleted scenarios. The best joint fit of transit and eclipse with near-solar water abundance and strong methane depletion matches both data points within 1σ (red, also see red curve in Figure 1). Consistency between the transit and eclipse is further underscored because the best fit model from the transit data alone (purple) captures the flux reversal between 3.6 μm and 4.5 μm correctly and presents a much better predictor of the eclipse data than our fiducial methane-rich self-consistent model (orange).

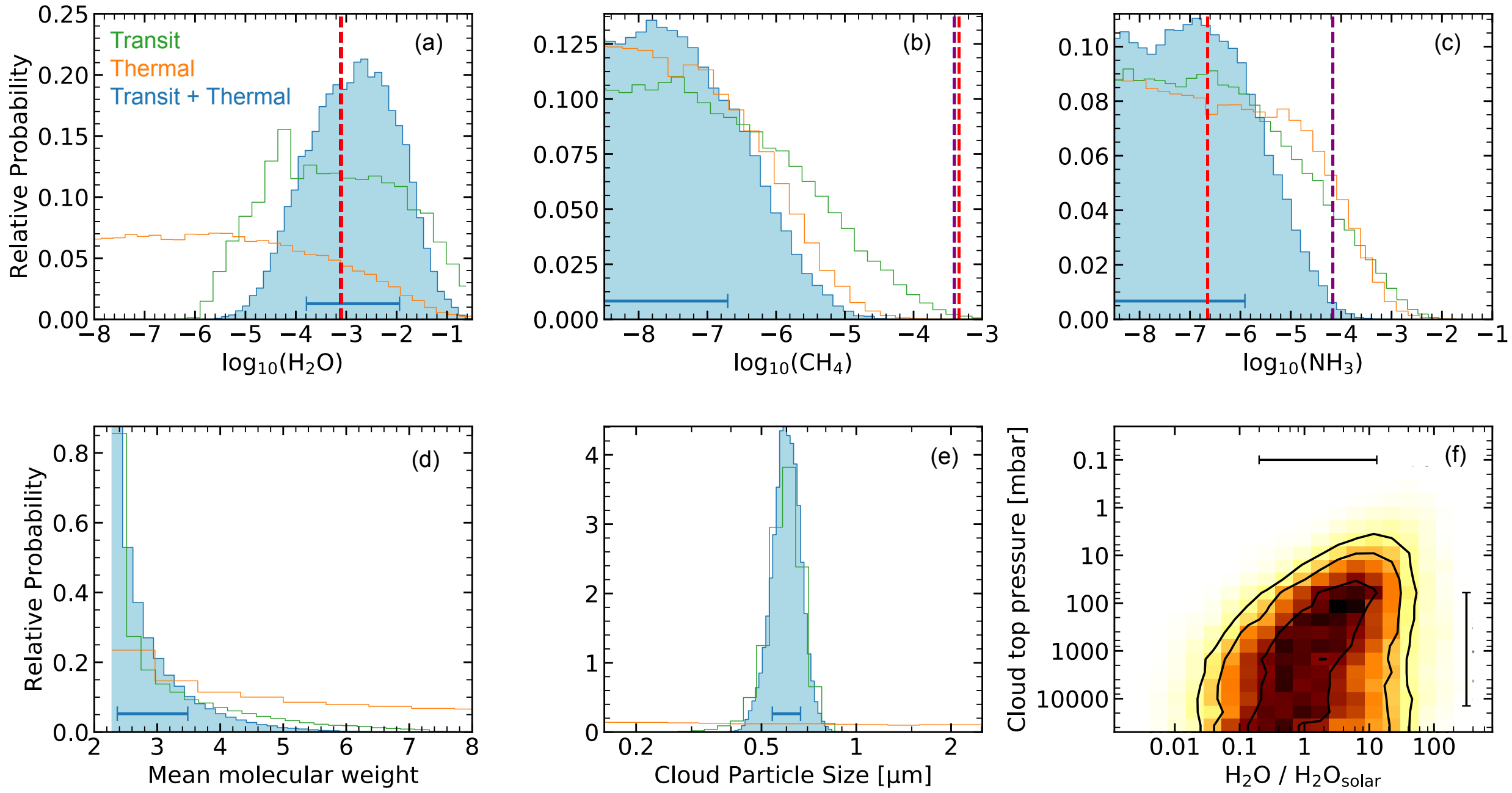

**Figure 4: Constraints on gas composition and cloud properties in the atmosphere of GJ 3470b.** The six panels show the marginalized probability distributions for the abundances of water vapor (a), methane (b), ammonia (c), the mean molecular mass (d), the cloud particle size (e), as well as the two-dimensional marginalized joint probability of the water abundance and the atmospheric pressure at which the clouds become optically thick (τ=1) (f). The blue 1D distributions indicate the posterior constraints derived from a joint analysis of all our transit and eclipse data. The blue horizontal bar indicates the central 68% probability interval. Green and orange show the equivalent distributions based on only the transit and only the eclipse data, respectively. For comparison, the vertical dashed lines indicate the molecular abundances for a 1 x solar metallicity atmosphere in chemical equilibrium (red) and based on the full photochemistry kinetics model (purple). Black lines in panel f indicate the 1σ , 2σ, and 3σ probability contours and the vertical and horizontal bars illustrate the 1σ uncertainties on the water vapour abundance and cloud top pressure individually. The water vapor abundance and mean molecular mass is found to be consistent with a gas giant-like atmosphere with a near-solar composition. Methane is found to be depleted by at least a factor of 30-50 at greater than 99.7% confidence. Ammonia is found to be depleted at >95% confidence compared to the expectation from chemical kinetics. Panel f shows the correlation between the water abundance and clouds due to their opposing effects on the strength of the 1.4μm water band in transit. The transit data also provide sharp constraints on the cloud particle size (panel e). A corner plot of all marginalized distributions is shown in Supplementary Figure 7.

# Methods

We jointly analyze all twelve HST and Spitzer transits taken between 2012 and 2017 (Supplementary Table 1) in our modular Exoplanet Transits Eclipses & Phasecurves (ExoTEP) framework. ExoTEP first reduces all raw data into standardized spectroscopic light curves and then determines the visible-to-IR transmission spectrum by computing the joint posterior probability of the transit and systematics model parameters of all instruments in one joint MCMC analysis. The 20 secondary eclipses per Spitzer channel are similarly fit in a global analysis of the planet's dayside emission data. Finally, we interpret GJ 3470b's transmission spectrum and the broadband Spitzer/IRAC secondary eclipse observations using a free atmospheric retrieval as well as a suite of self-consistent radiative/convective heat transfer and (dis-)equilibrium chemistry models. We performed a joint retrieval analysis of all data together as well as transits, eclipses, and subsets of the transit data individually. To minimize the sensitivity of our results to prior assumptions on the atmospheric processes, we focused on a "free" retrieval analysis with freely parameterized molecular abundances, a freely parameterized temperature structure, and freely parameterized properties of the upper cloud/haze deck.

## Observations and data reduction

### HST/WFC3 transits.

Each of three HST/WFC3 transit observations consisted of four 96-minute telescope orbits, with 48-min gaps in phase coverage between target visibility periods due to Earth occultation. The HST/WFC3 data were taken using the G141 grism and the spatial scan mode, which scans the telescope during the exposure and thereby moves the spectrum perpendicularly to the dispersion direction on the detector[4,35]. The spatial scanning enables longer exposures and reduces the instrumental overhead time compared to staring mode observations. We optimized the observations efficiency by performing spatial scans of maximum length across almost the full detector sub-array (256x256). We observed forward and backward scans to further reduce the instrumental overhead. In this way, we achieved an overall duty cycle of 73% for the bright star GJ 3470b, collecting more than ten times the number of photons per orbit than the previously obtained WFC3 stare mode observations of this star with an overall integration efficiency of 6.8%[9]. After combining all three transits, we collected 32 times more photons than this previous study, resulting in a five-fold improvement in the precision of our transit depth measurement in the WFC3 bandpass.

As a result of the spatial scans and the grism dispersion, the star light was spread over a near rectangular, but slightly trapezoidal patch on the detector. The trapezoidal shape of the illuminated patch on the detector results from the slight change in dispersion on the detector as the star moves along the detector's $y$-axis, leading to a 2–3-pixel shift in the $x$ direction for a given wavelength. As in previous studies[35], we minimize the sky background contribution in our full-frame images by co-adding the differences between consecutive up-the-ramp samples only for regions containing the target spectrum during the time between the consecutive nondestructive reads. We also flat-fielded the data using the wavelength-dependent flat field data provided by STScI and replace bad pixels by interpolating spatially.

To account for the trapezoidal shape, we first computed the 2D wavelength solution for spatial scans on the detector[36, 37]. We then compute lines of constant wavelength and define trapezoidal patches based on our predefined wavelength bins. We integrate over each of these trapezoidal patches for every time step to obtain the uncorrected spectrophotometric light curves. For each frame, we apply a small correction to the x position of the patch boundaries to account for the small drift of the star throughout the transit observation. Unlike previous studies[10,35,36], we do not pre-smooth the data in the dispersion direction. Instead, we integrate the flux across the trapezoidal patch by directly adding the pixel counts for all pixels that are fully within the trapezoidal patch, including fractions of the pixel counts for pixels that are intersected by the line of constant wavelength. The exact fraction added to the wavelength bin is determined by performing a local sub-pixel spline interpolation at the location of the intersected pixels and then adding fractions of the total counts proportional to the 2D integral of that spline to adjacent wavelength bins. This procedure ensures that the total flux is conserved. Combined with light curve fitting discussed below, this method delivers photon-noise limited residuals with no sign of contamination for both the white light and spectroscopic light curves (Supplementary Figures 1, 11, 12, 13).

## HST/STIS transits

We observed three transits of GJ 3470b using the G750L grism on HST/STIS to measure GJ 3470b's transmission spectrum between 0.55 and 1.0 μm. The STIS observations were taken in stare mode, i.e. without moving the telescope. The relatively small overhead for the read-out of our 1024x128 subarray resulted in a duty cycle of 89.0%. We performed the initial reduction of the raw data using the latest version of CALSTIS, and removed bad pixels and cosmic ray hits using a custom-made Python routine that searched for outliers in both position and time. For each image frame, we then extracted the 1D spectrum using spectral aperture extraction across a 25-pixel wide aperture. The initial wavelength calibration was provided by CALSTIS. We corrected for sub-pixel wavelength shifts in the dispersion direction over the course of the visit by finding the best fitting offset and amplitude between each 1D and the median 1D spectrum of the visit. We used the amplitudes from this optimization as the data points of the white light curve, as in previous studies [35]. Finally, we used the wavelength-corrected 1D spectra from each exposure to extract the flux value for 50-nm-wide spectral bandpasses, which we then combined to form the data points in the photometric light curves for each spectral channel.

## Spitzer/IRAC transits and eclipses

We analyzed a total of 6 Spitzer transits and 20 Spitzer eclipses of GJ 3470b in this work: 3 transits and 10 eclipses in each IRAC channel at 3.6 and 4.5 μm. Two of the 4.5μm Spitzer transits were previously published[38], the remaining 24 visits are previously unpublished. We used the standard peak-up pointing mode for the observations, which places the star reliably in the center of a pixel after allowing for an initial 30-minute settling time at the new pointing. We observed our target in subarray mode with 0.4 s exposures in both bandpasses, with a total duration of 6.44 hours (54,144 images) and 6.45 hours (54,272 images) of science observations for each 3.6 and 4.5 μm visit, respectively. Following standard procedure, we use the flat-fielded and dark-subtracted "Basic Calibrated Data" (BCD) images provided by the standard Spitzer pipeline for our analysis and estimated the sky background following Ref. [39]. We then determined the position of the star in each image using flux-weighted centroiding with a radius of 3.5 pixels, and calculate the flux in a circular aperture with radii of [2.0, 2.1, 2.2, 2.3, 2.4, 2.5, 2.6, 2.7, 2.8, 2.9, 3.0, 3.5, 4.0, 4.5, 5.0] pixels to create our photometric time series. We also consider an alternative version of the photometry utilizing a time-varying aperture, where we scale the radius of the aperture proportionally to the square root of the noise pixel parameter, which is proportional to the full width half max (FWHM) of the stellar point spread function [17,40,41]. As discussed in Refs. [39,42], we later optimize our choice of aperture, bin size, and trim duration individually for each visit by selecting the options which simultaneously minimize the RMS of the unbinned residuals as well as the time-correlated noise in the data in the individual fits. We quantify this time-correlated noise component by calculating the RMS as a function of bin size in steps of powers of two points per bin, and pick the best aperture, bin size, and trim duration that minimizes the least-squares difference between the actual RMS vs bin size and the white noise prediction for the individual visit.

# Transit white light curve fitting

We perform a global analysis of all *WFC3*, *STIS*, and *Spitzer* transit light curves by simultaneously fitting our transit light curve model, the instrument systematics models, and photometric noise parameters within one joint Markov Chain Monte Carlo analysis. Our analysis framework, ExoTEP, enables us to perform this analysis from extracted raw photometry to the transit parameters and their uncertainties as a single, statistically consistent Bayesian analysis. The main outputs of the analysis are the global transit parameters ($a/R_*$, b) as well as the planet-star radius ratios in the STIS bandpass (0.55-1.01µm), the WFC3 bandpass (1.1-1.7µm), and the two IRAC bandpasses (3.6 and 4.5 µm). To help the convergence within this high-dimensional parameter space, ExoTEP first fits each transit light curve individually and then automatically uses the best-fitting systematics model parameters from those individual fits as initial conditions in the global MCMC fit. We opt to fit GJ 3470b's ephemeris ($T_0$, P) with only the Spitzer transit observations because they cover the transits continuously at high cadence and are spaced over 5 years; hence, providing exquisite constraints on the GJ 3470b's transit ephemeris.

## HST/WFC3 Instrument Model

Our uncorrected WFC3 transit light curves exhibit well-documented systematic trends in flux with time, including visit-long slopes and orbit-long exponential ramps [4,35,43]. We account for these systematics by simultaneously fitting the data with the transit model and the analytical model-ramp function

$$S_{WFC3}(t) = (cd(t) + vt_v) \cdot (1 - \exp(-at_{orb} - b)) \qquad (1)$$

to correct for these instrumental systematics [26,43]. Here, $S_{WFC3}(t)$ is the transit model computed for the WFC3 bandpass, $c$ is a normalization constant, $d(t)$ is 1 for forward scans and $d$ for backward scans, $v$ is the visit-long linear slope, a and b are the rate constant and amplitude of the orbit-long exponential slope, and $t_v$ and $t_{orb}$ are the time in hours since start of the visit and the start of the observations within the current orbit. Following standard procedure, we discard the first HST orbit of each visit in the analysis because the amplitude of the ramp is larger than for the remaining orbits. We also remove the first forward and first backward scan exposures of each orbit, which improves the overall light curve fit.

## HST/STIS Instrument Model

Our STIS transit light curves exhibit ramp-like systematic trends comparable to those seen in previous STIS exoplanet data sets. STIS data show a visit-long slope and orbit-long ramps that result from the temperature settling of the telescope and the reinitiating of the read-out sequence at the beginning of each orbit. Following standard procedure[44–48], we correct for these systematics by simultaneously fitting the transit model and the analytical systematics model

$$S_{STIS}(t) = (c + vt_v) \cdot (1 + p_1 t_{orb} + p_2 t_{orb}^2 + p_3 t_{orb}^3 + p_4 t_{orb}^4) \qquad (2)$$

to the each STIS transit data set. Here, $c$ and $v$ are again the normalization constant and the visit-long linear slope, and $p_1$ to $p_4$ are coefficients to describe the systematic trend within each orbit via a $4^{th}$ order polynomial function. As in the analysis of WFC3, we discard the first HST orbit of each visit and remove the first two exposures of each orbit. We explored a wide range of more complex systematics models that simultaneously detrend against the x and y position and slope of the spectral trace on the detector as discussed in detail in Ref. [48], but found no substantial improvement in scatter or changes in our conclusions.

## Spitzer/IRAC Instrument Model

Our *Spitzer/IRAC* instrument model accounts for intra-pixel sensitivity variations and temporal sensitivity changes using the modified pixel-level decorrelation (PLD) model described in Ref [42]. Our instrument model is

$$S_{Spitzer}(t_i) = 1 + Ae^{-t_i/\tau} + mt + \frac{\sum_{k=1}^{9} w_k D_k(t_i)}{\sum_{k=1}^{9} D_k(t_i)} \;, \qquad (3)$$

where the systematics model $S_{Spitzer}(t_i)$ is composed of a fitted linear-exponential ramp in time ($Ae^{-t_i/\tau} + mt$) and the pixel-level decorrelation (PLD) term[49]. The $D_k(t_i)$'s in the PLD term are the raw counts in the 3x3 pixels, k=1…9, covering the central region of the PSF. In the numerator, these raw data values are multiplied by nine time-independent PLD weights, $\{w_1 \dots w_9\}$, fitted as free parameters in the light curve analysis. Together with the linear slope $m$, the instrument model therefore includes 10 free instrument fitting parameters to capture the intrapixel sensitivity variations and temporal sensitivity changes.

## Transit Model and MCMC Analysis

We compute the transit light curve $f(t_i)$ using the Batman implementation[50] of the standard transit equations[51]. In our joint white light curve fit, we allow for four different transit depths in the WFC3, STIS, IRAC 3.6µm, and IRAC 4.5µm bandpasses, but jointly fit a single set of transit geometry parameters (a/R*, b) to be consistent across the entire data set. The limb darkening coefficients are computed specifically for the STIS, WFC3, and Spitzer/IRAC bandpasses following the same procedure detailed in Ref. [48] using the PHOENIX stellar models[52] for a variety of different parameterizations[53]. Our fiducial stellar parameters for GJ 3470 are in T = 3652±50 K, log(g) = 4.843±0.035, and Z = 0.2±0.1 dex [34,54,55], and we perform the same wide sensitivity tests as detailed Ref. [48]. We account for the cadence of the observations by numerically integrating in time from the start to end of the individual exposures. We also account for the small but non-zero eccentricity independently constrained by the eclipse observations discussed below. Given the transit model and instrument sensitivity described above, we compute the log-likelihood function for the joint white light curve as

$$\ln \mathcal{L} = \sum_{V=1}^{N} \left[ -n_V \ln \sigma_V - \frac{n_V}{2} \ln 2\pi - \sum_{i=1}^{n_V} \frac{[D_V(t_i) - S_V(t_i) \cdot f_V(t_i)]^2}{\sigma_V^2} \right], \qquad , (4)$$

where the log likelihood is summed over all $N$ visits including the three WFC3, the three STIS, the six Spitzer/IRAC transits. The contribution from each visit is computed by summing the likelihood for each of the $n_V$ data points, $D_V(t_i)$, and simultaneously fitting a free photometric noise parameter $\sigma_V$ for each visit. We conservatively opt to fit for independent noise parameters for each visit to allow for visit-to-visit variations in the scatter in the data. The systematics model $S_V(t_i)$ is different for each visit and free parameters are included in the joint fit according to the systematics models (Equations 1,2, and 3) for the instrument used in the visit. Based on this likelihood function, we simultaneously compute the best estimates and joint posterior distribution of all astrophysical and systematics model parameters using the emcee package[56], a Python implementation of the Affine Invariant Markov Chain Monte Carlo (AI-MCMC) Ensemble sampler[57]. We assume flat priors on all parameters and initialize the global fit with the best estimates from MCMC fits to the individual transits. Fitted white light curves and their residuals for WFC3, STIS, and Spitzer are shown in Supplementary Figure 1, 2, and 3. We also perform individual fits to each of the transits. The transit depth measurements at each bandpass are highly repeatable and consistent within their uncertainties (Supplementary Figure 4). A detailed analysis of all the residuals are provided in the Supplementary Figures 9 – 17.

## Constraints on possible star spot effects

None of our twelve *HST* and *Spitzer* light curves show any direct evidence for star spot crossing events. The strongest constraints present the three HST/WFC3 white light curves which, with a Gaussian residual scatter of approximately 40 ppm per single data point, would have been highly sensitive to identifying the crossing of star spots along the planet chord [18]. In addition, we also find that all individual transit depth measurements taken over multiple epochs are consistent within their statistical uncertainties and with the joint fit (Supplementary Figure 4). Again, HST/WFC3 provides the strongest constraints and shows full consistency of the transit depths within 20 ppm. This consistency in time provides strong confidence that temporal variations of the star have no effect on the main conclusion derived in this work. Still, in what follows, we quantify the constraints using stellar models, and we account for possible common-mode effects in the retrieval analysis.

To calculate the potential effect of star spots on our results, we conservatively adopt a large spot coverage of 12.5-percent of the stellar disk and a spot temperature 250 K less than the photosphere, based on the values as reported for GJ 3470 in Ref. [13]. We then represent the stellar atmosphere inside and outside the star spots by two different Phoenix model atmospheres and allow for the possibility that some of the spots are occulted by the planet during transit and some are unocculted. Using this model, we investigate the potential effects of star spots by defining the fractional area of spots occulted during transit, $f$, and exploring the effects on the transmission spectrum across the whole range of $f$ values consistent with the observed repeatability of the WFC3 white light curve transit depth measurements (Supplementary Figure 4). For example, given the spot coverage of 12.5-percent, $f=0.1$ means that the spots occulted during transit cover 1.25-percent of the stellar disk and the unocculted spots cover 11.25-percent of the disk.

Supplementary Figure 4e shows the effect on the planet's inferred transmission spectrum for five representative values of $f$. For large values of $f$, a larger fraction of star spots is occulted during transit. As a result, less stellar flux is blocked by the transiting planet and the planet radius appears smaller, especially at the shortest wavelength (green and red lines). The opposite effect occurs for small values of f; unocculted spots cause the planet to appear larger, especially at the shortest wavelengths, because an overproportioned fraction of the stellar flux is blocked during transit (blue and purple curves). Finally, the effects become near zero (black curve) if the fraction of the spots that are occulted is equal to the fraction of the stellar disk crossed by the planet during the entire transit (unbiased crossing).

Quantitatively, we assess the effects of this departures from unbiased spot crossing event on the measured white light curve transit by integrating the star spectrum over the bandpass of the WFC3 instrument. We find that, for fractions f=0.06 and f=0.13, the star spot effects reach as high as 20 ppm in the WFC3 white light curve band pass, the maximum value to be consistent with the repeatability of the WFC3 white light curve measurements (Supplementary Figure 4a). In these limiting cases, the departures from unbiased spot crossing by the planet can produce effects on the planet's spectrum of maximum 150 ppm at STIS wavelengths, but no more than $\pm 15$ ppm in the amplitude of the WFC3 water absorption, and a few ppm at Spitzer wavelengths (Supplementary Figure 4e). However, even in these extreme cases, the effects of star spots on the WFC3 and Spitzer points are small compared to the uncertainties of the spectral transit depth measurement and the strength of the water detection. In addition, we find a consistent three transit measurements across our three STIS visits, and we deem it to be highly unlikely that a discordant STIS spectrum of the planet was fortuitously combined with a star spot effect to produce the consistent transit radius that we observe. We still include these common-mode uncertainties as nuisance parameters in the retrieval as discussed below.

## Transit spectroscopy

The ramp-like instrument systematics in our WFC3 observations are, to first order, independent of wavelength and position on the detector. Following standard practice, we therefore remove the ramp-like systematics from the photometric time series of each spectroscopic channel by applying corrections based on the white light systematics model. We apply this pre-correction to the spectroscopic time series in two different ways and find negligible differences in the resulting transit depth fits. The first approach is to divide the spectroscopic time series by the systematics models given in Eq. (1) using the best-fitting parameters from the white light curve fit [4]. The second approach is to divide the spectroscopic time series by the ratio of the uncorrected white light time series and the best fitting white-light transit model. The latter approach mathematically resembles the approach of fitting white-light-divided time series taken in Ref. [35]. The advantage of both approaches is that the spectroscopic time series are pre-corrected using a systematics model derived from high SNR broadband light curves allowing for fewer systematics parameters to be fit to spectroscopic light curves. Finally, we obtain the transit depth for each of the spectroscopic channels by performing joint MCMC fits to the pre-corrected spectroscopic times series from all three WFC3 transits for that spectroscopic channel. Similar to the white light curve fit, we simultaneously fit a transit model and a systematics model using MCMC. The only differences are that the parameters ($a/R_*$, b, $T_0$, P) are fixed to the best-fitting values from the white light curve fit and that systematics model can be highly simplified because the dominant time-dependent ramp have already been removed.

We obtain the STIS transmission spectrum by performing independent fits to the times series of eight 50-nm-wide spectroscopic channels between 0.55 and 0.95 µm. As with WFC3, we fit the transit depth for each of the spectroscopic channels by performing a joint MCMC fit to all three STIS transits. Following standard practice, however, we use the full STIS systematics model (Eq. 2) for each of our channels [46,58]. We explored pre-correcting the spectroscopic light curves with the white light curve, but found that the systematics are not sufficiently uniform across the spectroscopic channels to benefit from this approach. The final transmission spectrum is shown in Figure 2 and the light curve fit for a typical spectroscopic channel are shown in Supplementary Figure 2. Within the uncertainties, all STIS transit depths are consistent with a single value and with the WFC3 data, supporting the conclusion from WFC3 that high altitude clouds are present in the atmosphere along the terminator. We find no evidence for the Rayleigh scattering slope reported by [12] within the wavelength range covered by our observations (> 0.55 µm). The transit depth uncertainties in the STIS are larger than the ones derived from the WFC3 data because the instrument throughput of STIS is three times lower and the M2 star GJ3470b is approximately 40 fainter in the STIS bandpass than in the WFC3 bandpass.

## Secondary eclipse analysis

As for the transits, we fit each Spitzer eclipse time series with the pixel-level decorrelation (PLD) model[42,49]. We include both a linear and (for the 3.6 μm data only) an exponential function of time. Using fits to each individual eclipse, we first identify the optimal choice of aperture, bin size, and trim duration which minimizes simultaneously minimize the RMS of the unbinned residuals as well as the time-correlated noise in the data (Supplementary Table 2).

As expected in the Spitzer program design, we do not detect the eclipse with greater than 3σ significance in any individual visit, and therefore carry out a simultaneous fit to the ten eclipse light curves in each bandpass. We detect the eclipse at 3.6 micron with 4.7 sigma significance and place an upper limit on the 4.5 micron eclipse depth. The constraints on the eclipse depths and brightness temperatures are $F_P/F_* = 115^{+27}_{-26}$ and $T_\mathrm{B} = 844^{+33}_{-39}$ at 3.6 μm and $F_P/F_* = 3 \pm 22$ and $T_\mathrm{B} < 532$ K at 4.5 μm, respectively. For the global fit we allow all of the instrumental noise model parameters to vary independently for each visit, but assume a common depth and orbital phase for the secondary eclipse signal. The eclipse is slightly offset relative to the estimated eclipse time for a perfectly circular orbit. This is consistent with the radial velocity of GJ 3470b, which independently confirm a small, but non-zero eccentricity [55]. The resulting eclipse depths for the global and individual eclipse fits are shown in Supplementary Figure 5. Here, we fixed the eclipse time to the value of the global fit at 3.6μm. Consistent with the uncertainties, the ten individual measurements are randomly distributed around the global fit. Six of ten and seven of ten data points are within the 68% confidence interval for the observations at 3.6 and 4.5 μm, respectively.

We note that there are either eleven (4.5 μm) or thirteen (3.6 μm) free parameters for the instrumental noise model in the individual eclipse fits, including nine pixel-level light curve coefficients, a linear function of time, the measurement error, and a timescale and a normalization term for the exponential function (3.6 μm data only). This results in a prohibitively large global model, which includes 132 free parameters in the case of the 3.6 μm data. We instead elect to reduce the degrees of freedom in our model by using linear regression to determine the nine best-fit pixel-level light curve coefficients at each step in our MCMC fit. This reduces the total number of free parameters to 42 and 22 for the 3.6 and 4.5 μm fits, respectively. While this approach is not formally correct, as we are effectively optimizing a subset of our model parameters rather than marginalizing (i.e., integrating the posterior probability distribution over all possible parameter values), we find that when comparing fits to individual eclipse observations using linear regression to those utilizing the full PLD model the reported uncertainty on the eclipse depth decreases by less than 10%.

## Atmospheric retrieval

We interpret GJ 3470b's transmission and thermal spectra using a suite of established modeling tools, including atmospheric retrieval and thermo- and photochemical kinetics and transport modeling. As we combined a comprehensive data set of transit and eclipse data covering 0.55 to 5 μm, our analysis provides substantially narrower constraints on the composition than previously reported[12,13,59,60]. On the atmospheric retrieval side, we use a derivative of the SCARLET atmospheric retrieval suite[4,5,11,61–63] to interpret the transit and eclipse spectra jointly and individually. The SCARLET forward model computes star light transmitted through the atmosphere as well as the upwelling disk-integrated emission given a set of molecular abundances, the temperature structure, and cloud properties. The forward model is then coupled to the Affine Invariant Markov chain Monte Carlo (MCMC) Ensemble sampler emcee[56] to perform the estimation of the gas composition, thermal, and cloud parameters. We retrieve the joint posterior probability distribution of the volume mixing ratios of 6 molecular gases ($H_2O$, $CH_4$, CO, $CO_2$, $NH_3$, and HCN), 5 parameters for the temperature structure[64,65], 3 parameters for Mie-scattering cloud particles, and 2 nuisance parameters for common-mode uncertainties within the transit depths measurements from WFC3 and STIS. Our Mie scattering cloud parameterization is newly developed while our parameterization of the temperature structure and the molecular abundances are common practice.

**Mie scattering cloud parameterization.** Since the data show direct evidence for scattering by finite-sized particles, we developed a new Mie-scattering cloud parameterization for atmospheric retrieval that enables us to explore the properties of the cloud particles near the top of the clouds. Mie scattering had previously been widely used in atmospheric forward models of exoplanets[11,14,17,66], but not directly within a Bayesian atmospheric retrieval analysis. Driven by the observational geometry, the objective of the new parameterization is to capture the wavelength dependent line-of-sight extinction of a wide range of cloud tops, while remaining low complexity and independent of preconceived ideas of cloud formation in this previously unexplored planet regime. We simultaneously accomplish these objectives by describing the clouds using three free parameters: the effective particle size near the cloud top ($R_{\text{part}}$), the scale height or 'fuzziness' of the clouds near the cloud top in units of the local gas density scale height ($H_{\text{part}}/H_{\text{gas}}$), and the atmospheric pressure at which the clouds become optically thick to grazing light beams at 1.5μm ($P_{\tau=1}$). For each set of the three cloud parameters, SCARLET then iteratively derives an altitude-dependent particle number density profile $n_{\text{part}}(z)$ near the cloud top to match the value for $P_{\tau=1}$, the pressure at which the optical depth for grazing light beams reaches $\tau = 1$. The particle size distribution is a log-normal particle size distribution centered around $R_{\text{part}}$ in each layer, and a full wavelength-dependent Mie scattering computation is performed for each particle size in the log-normal distribution to model the radiative effects of the finite-sized particles. This process is repeated for each parameter set suggested by the MCMC. For the wavelength-dependent refractive index of the clouds material, we assume the ones of KCl, and we find that our results are not substantial changed when replacing KCl by ZnS, $Na_2S$ or soots, which are also possible compositions for clouds on GJ 3470b[67,68] The described three parameter were chosen because they describe the line-of-sight extinction properties of the cloud top in a highly orthogonal way, ideal for the retrieval. The parameter $P_{\tau=1}$ predominately sets the vertical offset between the cloud extinction and the molecular absorption in the transmission spectrum, $R_{\text{part}}$ sets the wavelength-dependency of cloud opacities, and $H_{\text{part}}/H_{\text{gas}}$ predominately sets the amplitude of the extinction contrast across the transmission spectrum. We deliberately leave out parameters for the cloud base and lower clouds because they would not be constrained by the data. In the limit of large cloud particles, the models returns to the assumptions of gray clouds as applied in previous retrieval analyses[4,11,26,69]. Similarly in the limit of small particles, the new parameterization returns to the assumption of Rayleigh scattering hazes ($\sigma \propto \lambda^{-4}$).

**Parameterization of temperature profile.** The temperature–pressure profile is parameterized with five free parameters using the analytic approximation from Refs. [64,65]. The free parameterization is flexible enough to permit a wide range of thermal structures including thermal inversions, while resulting in smooth, physically motivated profiles consistent with radiative energy balance as the analytic approximation enforces radiative equilibrium in its derivation. In the three retrievals depicted in Figure 4, we allowed for fully independent temperature profiles near the terminator (transit; green) and on the dayside (secondary eclipse; orange). Only for the joint fit (blue), we assumed that the temperatures are approximatively similar. We do this because additional five parameters for the terminator temperature structure is computationally extremely expensive and not justified given that our transmission spectrum contains virtually no information on the temperature structure[11] and the temperature structure near the terminator is not the main focus of this work. Hence, the retrieved temperature structure and its uncertainties depicted Supplementary Figure 6 are dominated by the information in the secondary eclipse measurements. The increase in temperature from 0.01 bars to 10 bars is constrained by the observed brightness temperature contrast between the 3.6μm and 4.5μm Spitzer band passes. Still, we additionally verify that the assumption of similar temperature structure does not significantly affect the conclusions from our joint retrieval by running independent retrievals based on only the transmission spectrum and based on transit and eclipse, where we deliberately offset the temperatures by ±20%. No substantial changes in the retrieval results are observed because free retrievals that directly specify molecular abundances rather than modeling the chemistry are not very sensitive to the exact temperature structure [11].

**Parameterization of molecular composition.** Simultaneous with the cloud and temperature parameters, we retrieve six thermochemically plausible gases that absorb over the wavelengths covered by the observations ($H_2O$, $CH_4$, CO, $CO_2$, $NH_3$, and HCN). We assign uniform-in-log mixing ratio priors spanning from −10 to 0. Molecular hydrogen and helium (in solar proportions) are assumed to comprise the remaining gas such that all species sum to unity. We use the absorption cross-section from ExoMol for $CH_4$, $NH_3$, and HCN and the empirical opacities list from HITEMP for $H_2O$, CO, and $CO_2$ [70,71]. In the free retrieval, we make no a priori assumptions on the chemistry of our retrieved abundances.

**Nuisance parameters for common-mode uncertainties.** The analysis of the spectral light curves from WFC3 uses common mode corrections derived from the white light curve analysis to increase the relative precision and reduce the complexity of the systematics models (see above). We account for the small common-mode offsets that this can introduce between the WFC3 points and the other instruments by including a free offset nuisance parameter in the retrieval. Similarly, we introduce a white light curve offset nuisance parameter for STIS to account for the uncertainty introduced by the possible presence of unocculted star spots. The variance for the WFC3 offset prior is the WFC3 white light curve uncertainty (5ppm) in square sum of with the constraints from the star spot assessment (see star spot section). For STIS we only account for the stellar effects because no common mode correction is used in the analysis.

**Significance of molecule detection.** We determine the significance of the water detection based on full Bayesian evidence calculations as developed in Ref [11] and now widely applied, e.g. Refs. [26,72–74]. A molecule is deemed "detected" if the improvement in likelihoods from the addition of that given parameter outweighs the increase in prior volume. This is the most direct Bayesian approach to assess the significance of a detection, hence removing the need for the approximations in the commonly used Bayesian information criterion (BIC) as an estimator for significance. Here, we determine the significance of the water detection by removing water from the model and rerunning the retrieval to compute the new Bayesian evidence without water. We find the evidence ratio of the models with and without water, i.e. the Bayes factor, to be $B_{H2O}$=124,700 corresponding to a highly robust detection. For reference, a 5.2σ significance would be equivalent in the frequentist's framework. The lower Bayesian evidence for the model without water (one parameter less) is a direct result of the difference in maximum likelihood ($\Delta\chi^2_{\mathrm{min}} = 23.8$) as the model without water cannot match the WFC3 data.

## Self-consistent atmosphere models and photochemistry calculations

To put derived retrieval results and observations in context, we use self-consistent forward modeling tools to derive the expected molecular abundances and spectra for planetary atmospheres in thermal and chemical equilibrium (Figure 2, 3 and 4). A set of fiducial atmospheres was computed using the implementation described in Refs. [11,63]. Both models iteratively solve the radiative-convective heat transport and chemical equilibrium and result in virtually identical abundances and spectra. To investigate the deficiency in methane and ammonia, we then also apply the thermo- and photochemical kinetics and transport model[16,75,76] to assess the effects of non-equilibrium chemistry. The kinetics and transport model captures the three main chemical processes in planetary atmospheres: thermochemical equilibrium in the deep atmosphere, transport-induced quenching in the mid-atmosphere, and photochemistry in the upper atmosphere. A state-of-the-art reaction list of 1760 chemical reactions for 92 molecular species formed by the elements H, C, O, and N is adopted[16]. The temperature structure for the kinetics and transport simulations is set to the ones derived in the self-consistent forward models. The results from both the chemical equilibrium and the chemical kinetics/photochemistry calculations are indicated in Figure 3, demonstrating that the lack of methane ($CH_4$) cannot be explained by currently captured non-equilibrium effects in our 1D models. Photochemical destruction is expected to result in the reduction of the methane abundance in the uppermost atmosphere of GJ 3470b ($p < 10$ μbar), but the pressure levels that dominate the transmission spectrum are virtually unaffected by the methane destruction—as previously also shown in a theoretical modeling investigation of the sub-Neptune GJ 1214b[15] and GJ 436b[16,21] (Supplementary Figure 8).

**Data Availability.** The data presented in this work are publicly available in the Mikulski Archive for Space Telescope (MAST) and the Spitzer Heritage Archive (SHA).

Supplementary Information

# A Sub-Neptune Exoplanet with a Low-Metallicity Methane-Depleted Atmosphere and Mie-Scattering Clouds

BJÖRN BENNEKE[1], HEATHER A. KNUTSON[2], JOSHUA LOTHRINGER[3], IAN CROSSFIELD[4], JULIANNE MOSES[5], CAROLINE MORLEY[6], LAURA KREIDBERG[7], BJ FULTON[2], DIANA DRAGOMIR[4,17], ANDREW HOWARD[8], IAN WONG[9], JEAN-MICHEL DÉSERT[10], P.R. MCCULLOUGH[11], ELIZA M.-R. KEMPTON[12,13,], JONATHAN FORTNEY[14], RONALD GILLILAND[15], DRAKE DEMING[12], JOSHUA KAMMER[16]

[1] Department of Physics and Institute for Research on Exoplanets, Universit√© de Montr√©al, Montr√©al, QC, Canada
[2] Division of Geological and Planetary Sciences, California Institute of Technology, Pasadena, CA 91125, USA
[3] Lunar & Planetary Laboratory, University of Arizona, 1629 E. University Boulevard., Tucson, AZ, USA
[4] Department of Physics and Kavli Institute of Astronomy, Massachusetts Institute of Technology, 77 Massachusetts Ave, Cambridge, MA, 02139, USA
[5] Space Science Institute, 4750 Walnut Street, Suite 205, Boulder, CO 80301, USA
[6] Department of Astronomy, University of Texas, Austin, TX 78712, USA
[7] Department of Astronomy, Harvard University, 60 Garden Street, Cambridge, MA 02138, USA
[8] Department of Astronomy, California Institute of Technology, Pasadena, CA 91125, USA
[9] Department of Earth, Atmospheric, and Planetary Sciences, Massachusetts Institute of Technology, 77 Massachusetts Ave, Cambridge, MA, 02139, USA
[10] Anton Pannekoek Institute for Astronomy, University of Amsterdam, 1090 GE Amsterdam, The Netherlands
[11] Department of Physics and Astronomy, Johns Hopkins University, Baltimore, MD 21218, USA
[12] Department of Astronomy, University of Maryland, College Park, MD 20742, USA
[13] Department of Physics, Grinnell College, 1116 8th Avenue, Grinnell, IA 50112, USA
[14] Department of Astronomy, University of California, Santa Cruz, CA 95064, USA
[15] Space Telescope Science Institute, 3700 San Martin Dr., Baltimore, MD 21218, USA
[16] Southwest Research Institute, San Antonio TX, USA
[17] IPAC-NASA Exoplanet Science Institute Pasadena, CA 91125, USA
[18] NASA Hubble Fellow

| Instrument | Filter/Grism | Transit/Eclipse | Wavelength [µm] | UT Start Date |
|---|---|---|---|---|
| HST/STIS | G750L | Transits | 0.55 – 1.0 | 2015 Feb 07<br>2015 May 12<br>2016 Apr 03 |
| HST/WFC3 | G141 | Transits | 1.1 – 1.7 | 2015 Jan 28<br>2015 Mar 13<br>2015 Oct 22 |
| Spitzer/IRAC | Channel 1 | Transits | 3.0 – 4.0 | 2012 Dec 22<br>2017 Jan 25<br>2017 Feb 20 |
| Spitzer/IRAC | Channel 2 | Transits | 4.0 – 5.0 | 2012 Jun 11<br>2012 Jun 15<br>2013 Jan 01 |
| Spitzer/IRAC | Channel 1 | Eclipse | 3.0 – 4.0 | 2014 Jan 15<br>2014 Jan 28<br>2014 Jun 14<br>2014 Jun 24<br>2015 Jan 30<br>2015 Feb 06<br>2015 Feb 09<br>2015 Feb 12<br>2015 Jun 19<br>2015 Jul 19 |
| Spitzer/IRAC | Channel 2 | Eclipse | 4.0 – 5.0 | 2014 Jan 21<br>2014 Feb 4<br>2014 Jun 21<br>2014 Jul 10<br>2015 Jan 10<br>2015 Jan 13<br>2015 Jan 17<br>2015 Jan 20<br>2015 Jan 23<br>2015 Jan 27 |

Supplementary Table 1: Summary of presented transit and eclipse observation of GJ 3470b.

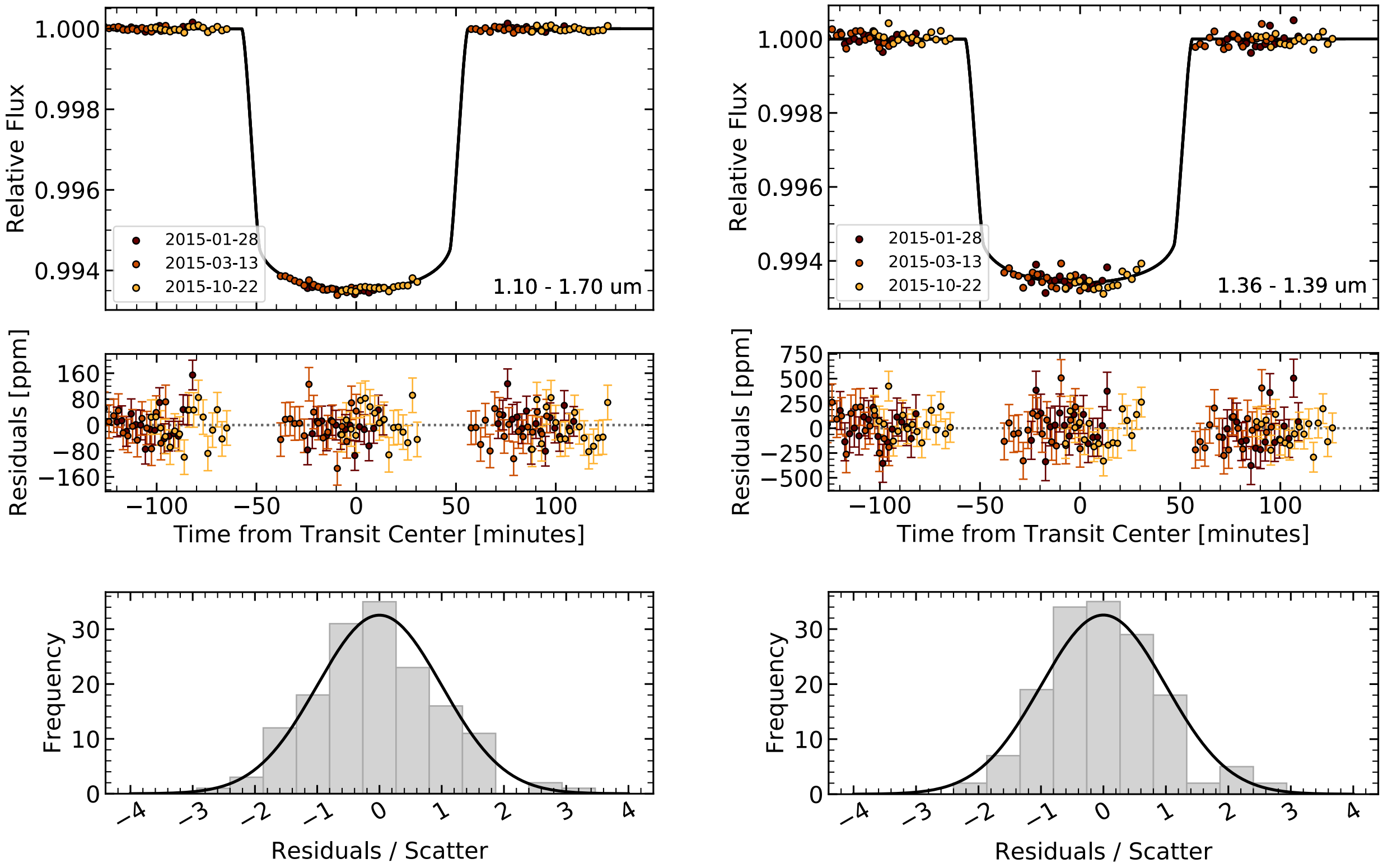

Supplementary Figure 1: White light curve fit (left) and a typical spectral light curve fit (right) from the joint analysis of the three WFC3 transit observations of GJ 3470b. The top panel shows the best fitting model light curves (black curve), overlaid with the systematics-corrected data (circles). Residuals from the light curve fits are shown in the middle panels. All corrected WFC3 light curve fits are free of obvious systematics. The bottom panels shows a histogram of the residuals normalized by the fitted photometric scatter parameter for each respective transit. The residuals follow the expected Gaussian distribution for photon noise limited observations.

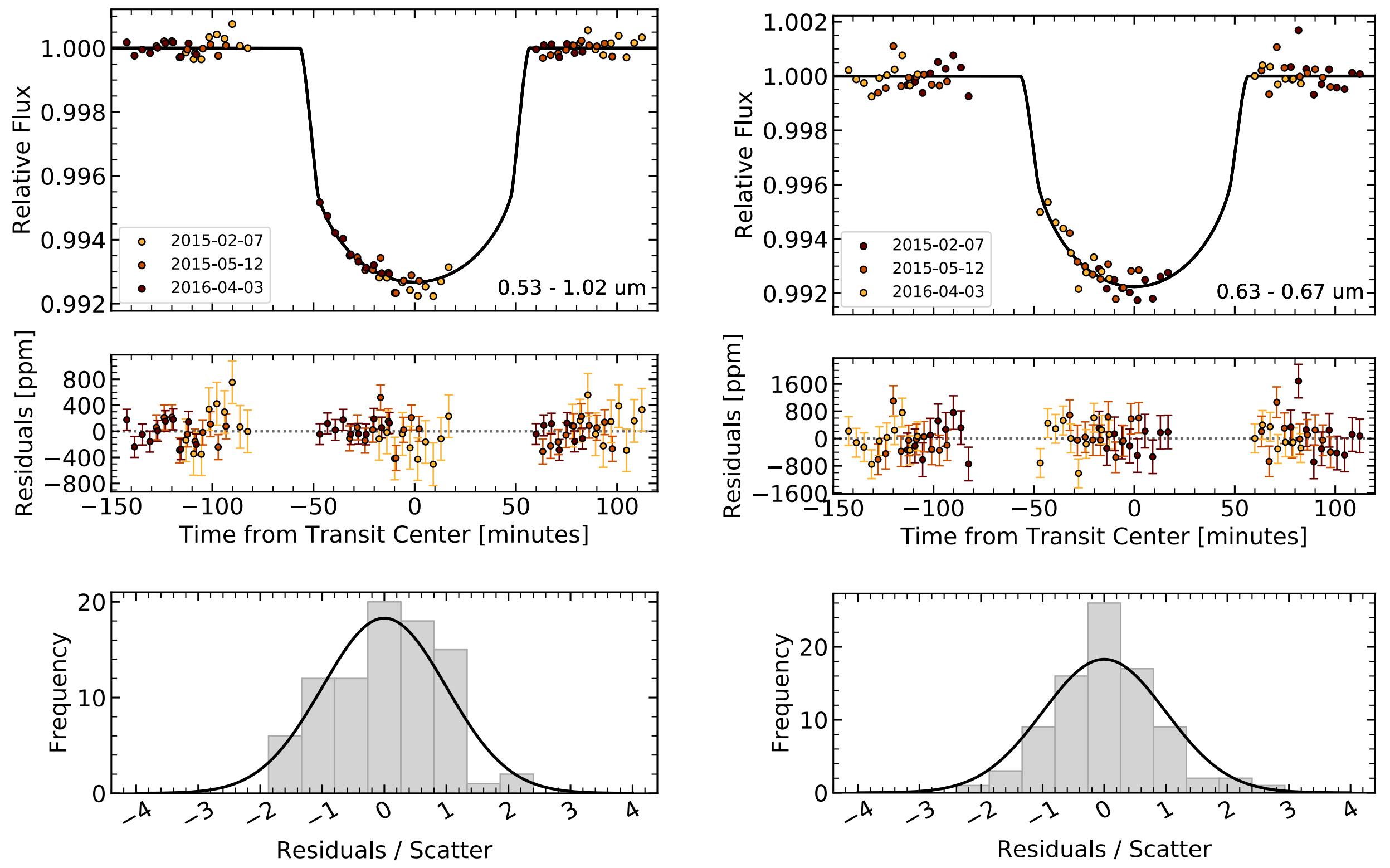

Supplementary Figure 2: White light curve fit (left) and a typical spectral light curve fit (right) from the joint analysis of the three STIS transit observations of GJ 3470b. The top panel shows the best fitting model light curves (black curve), overlaid with the systematics-corrected data (circles). Residuals from the light curve fits are shown in the middle panels. The bottom panels shows a histogram of the residuals normalized by the fitted photometric scatter parameter for each respective transit.

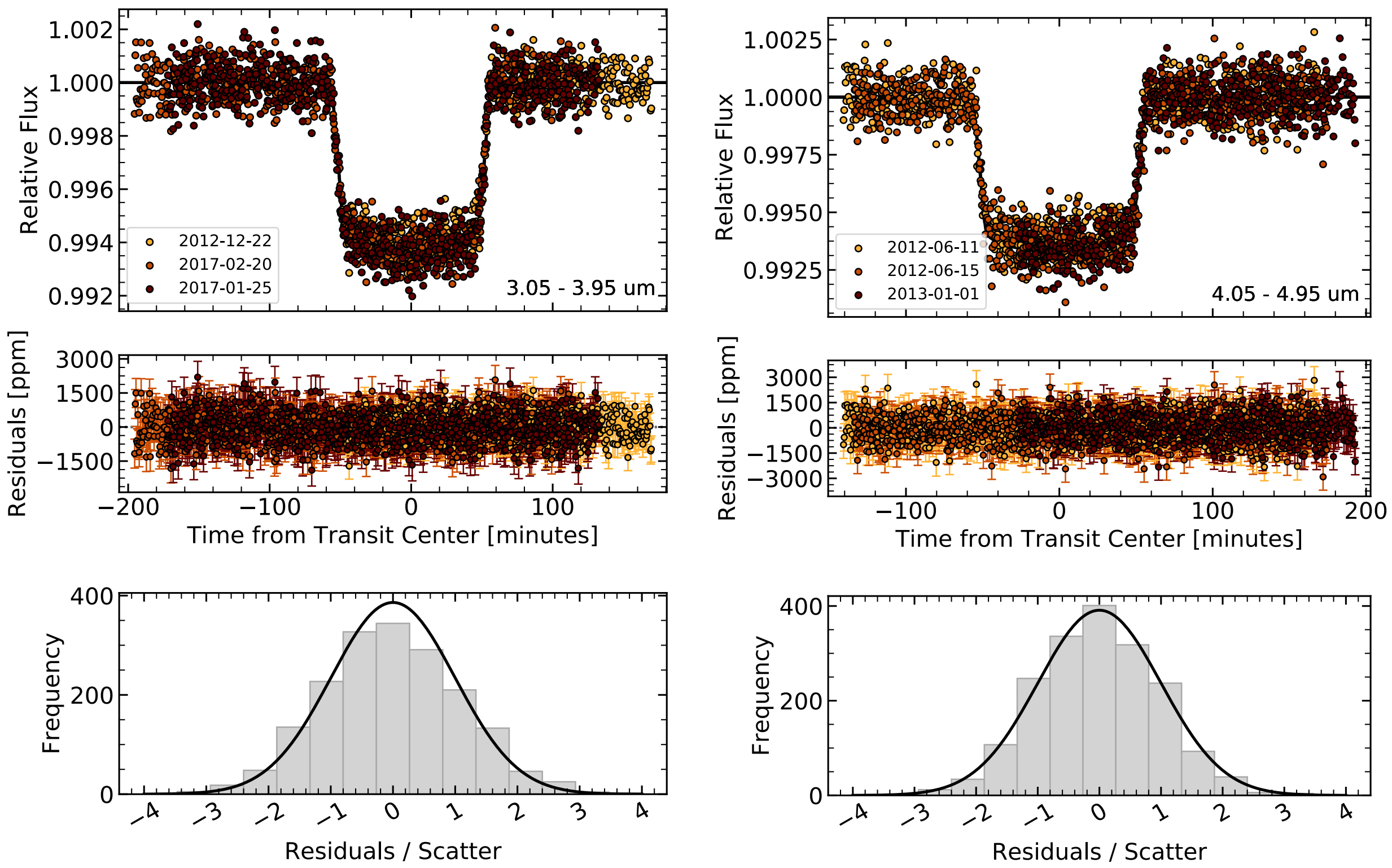

Supplementary Figure 3: Spitzer light curve fits of three 3.6µm transit (left) and three 4.5µm transits (right). The top panel shows the best fitting model light curves (black curve), overlaid with the systematics-corrected data (colored circles). Residuals from the light curve fits are shown in the middle panels. All corrected Spitzer light curve fits are free of obvious systematics. The bottom panels shows a histogram of the residuals normalized by the fitted photometric scatter parameter for each respective transit. The residuals follow the expected Gaussian distribution for photon noise limited observations.

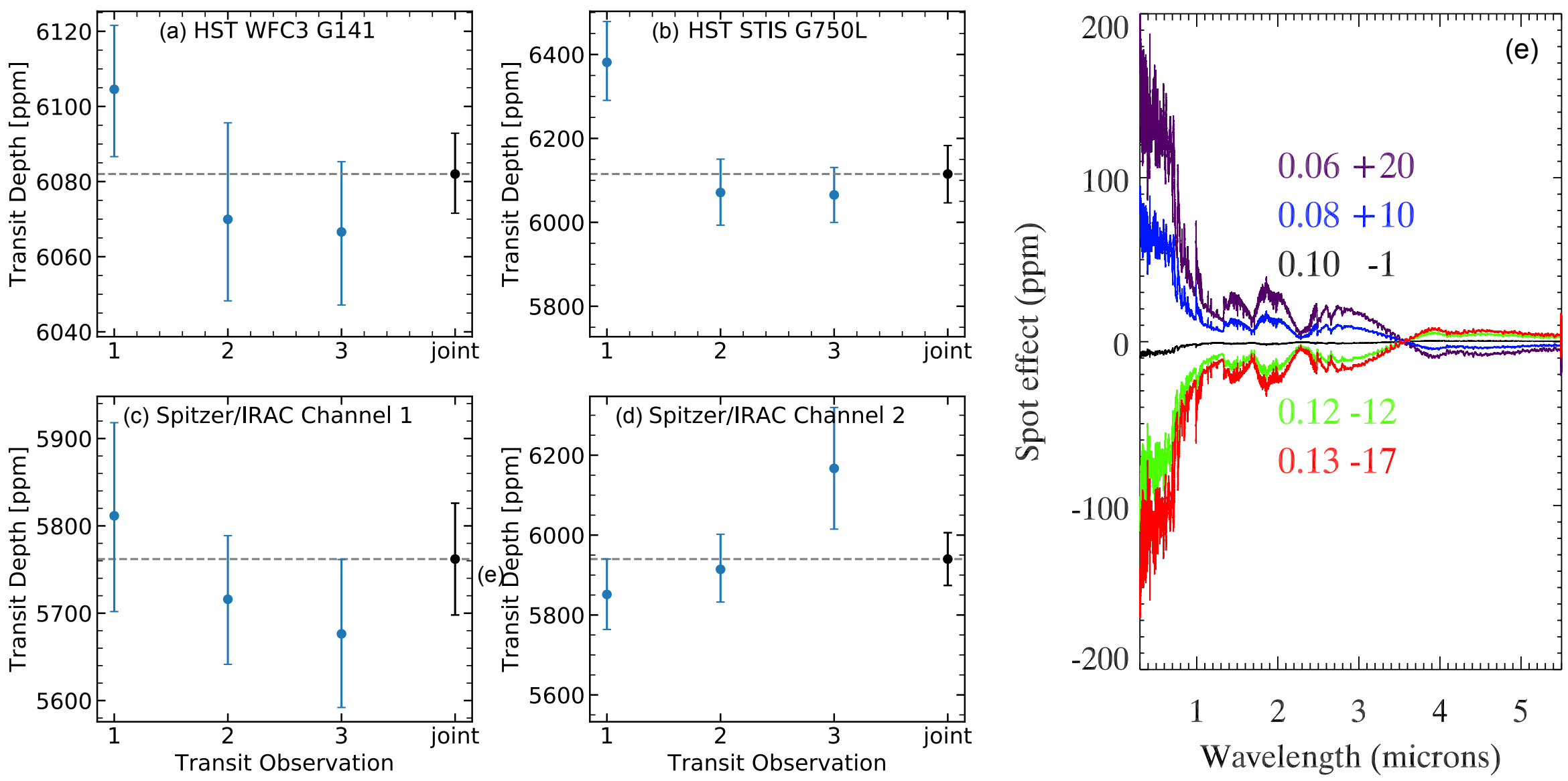

Supplementary Figure 4: (Left and center) Repeatability of transit depth measurements. Panels (a)-(d) show the transit depths from the individual transit fits (blue) and joint fit (black) for HST/WFC3 (a), HST/STIS (b), Spitzer/IRAC 3.6µm (c) and Spitzer/IRAC 4.5µm (d). The individual transit depth measurement are consistent over time within their statistical uncertainties and with the joint fit. (Right) Modeled spectra showing the potential effects of star spots on the apparent transmission spectrum. Colored curves indicate the effect for a fraction f of 0.06 (purple), 0.08 (blue), 0.1 (black), 0.12 (green) and 0.13 (red) as discussed in the Methods Section. The second value indicates the increase in the apparent transit depth within the WFC3 bandpass in parts-per-million (p.p.m).

| Visit # | $\lambda$ (μm) | UT Start Date | $t_{trim}$ (hr)[a] | $n_{bin}$ [a] | Varying aperture? | Noise Pixel Scaling | Average $r_{phot}$ [a] | Bkd (%)[b] |
|---|---|---|---|---|---|---|---|---|
| 1 | 3.6 | UT 2012 Dec 22 | 0.5 | 128 | no | | 2.7 | 0.91 |
| 2 | 3.6 | UT 2017 Jan 25 | 0.75 | 64 | no | | 2.8 | 1.01 |
| 3 | 3.6 | UT 2017 Feb 20 | 0.75 | 128 | no | | 2.5 | 1.04 |
| 1 | 4.5 | UT 2012 Jun 11 | 0.75 | 128 | no | | 2.6 | 0.29 |
| 2 | 4.5 | UT 2012 Jun 15 | 0.75 | 64 | no | | 2.7 | 0.32 |
| 3 | 4.5 | UT 2013 Jan 01 | 0.5 | 64 | no | | 2.7 | 0.35 |
| 1 | 3.6 | UT 2014 Jan 15 | 1.5 | 64 | yes | 1.1x scaling | 2.6 | 0.92 |
| 2 | 3.6 | UT 2014 Jan 28 | 0.5 | 192 | no | | 2.3 | 0.82 |
| 3 | 3.6 | UT 2014 Jun 14 | 0.5 | 128 | no | | 2.2 | 0.65 |
| 4 | 3.6 | UT 2014 Jun 24 | 1.5 | 128 | yes | 0.9x scaling | 2.5 | 0.67 |
| 5 | 3.6 | UT 2015 Jan 30 | 1.5 | 64 | no | | 2.2 | 0.69 |
| 6 | 3.6 | UT 2015 Feb 06 | 1.5 | 128 | yes | 1.2x scaling | 3.0 | 1.26 |
| 7 | 3.6 | UT 2015 Feb 09 | 0.5 | 192 | no | | 2.7 | 1.16 |
| 8 | 3.6 | UT 2015 Feb 12 | 1.5 | 192 | no | | 2.3 | 0.91 |
| 9 | 3.6 | UT 2015 Jun 19 | 0.5 | 128 | no | | 2.2 | 0.71 |
| 10 | 3.6 | UT 2015 Jul 19 | 0.5 | 192 | no | | 2.3 | 0.80 |
| 1 | 4.5 | UT 2014 Jan 21 | 1.5 | 64 | no | | 2.5 | 0.30 |
| 2 | 4.5 | UT 2014 Feb 04 | 1.0 | 128 | no | | 2.3 | 0.22 |
| 3 | 4.5 | UT 2014 Jun 21 | 1.0 | 64 | no | | 2.3 | 0.10 |
| 4 | 4.5 | UT 2014 Jul 10 | 1.0 | 64 | no | | 2.3 | 0.29 |
| 5 | 4.5 | UT 2015 Jan 10 | 1.5 | 64 | no | | 2.1 | 0.37 |
| 6 | 4.5 | UT 2015 Jan 13 | 1.0 | 128 | no | | 2.3 | 0.34 |
| 7 | 4.5 | UT 2015 Jan 17 | 1.5 | 64 | no | | 2.1 | 0.30 |
| 8 | 4.5 | UT 2015 Jan 20 | 1.0 | 128 | no | | 2.4 | 0.35 |
| 9 | 4.5 | UT 2015 Jan 23 | 0.5 | 128 | no | | 2.3 | 0.28 |
| 10 | 4.5 | UT 2015 Jan 27 | 1.5 | 64 | no | | 2.7 | 0.34 |

Supplementary Table 2: Summary of *Spitzer* transit observations (top) and eclipse observation (bottom)

[a] $t_{trim}$ is the amount of time trimmed from the start of each time series, $n_{bin}$ is the bin size used in the photometric fits, and $r_{phot}$ is the radius of the photometric aperture in pixels.

[b] Relative sky background contribution to the total flux in the selected aperture

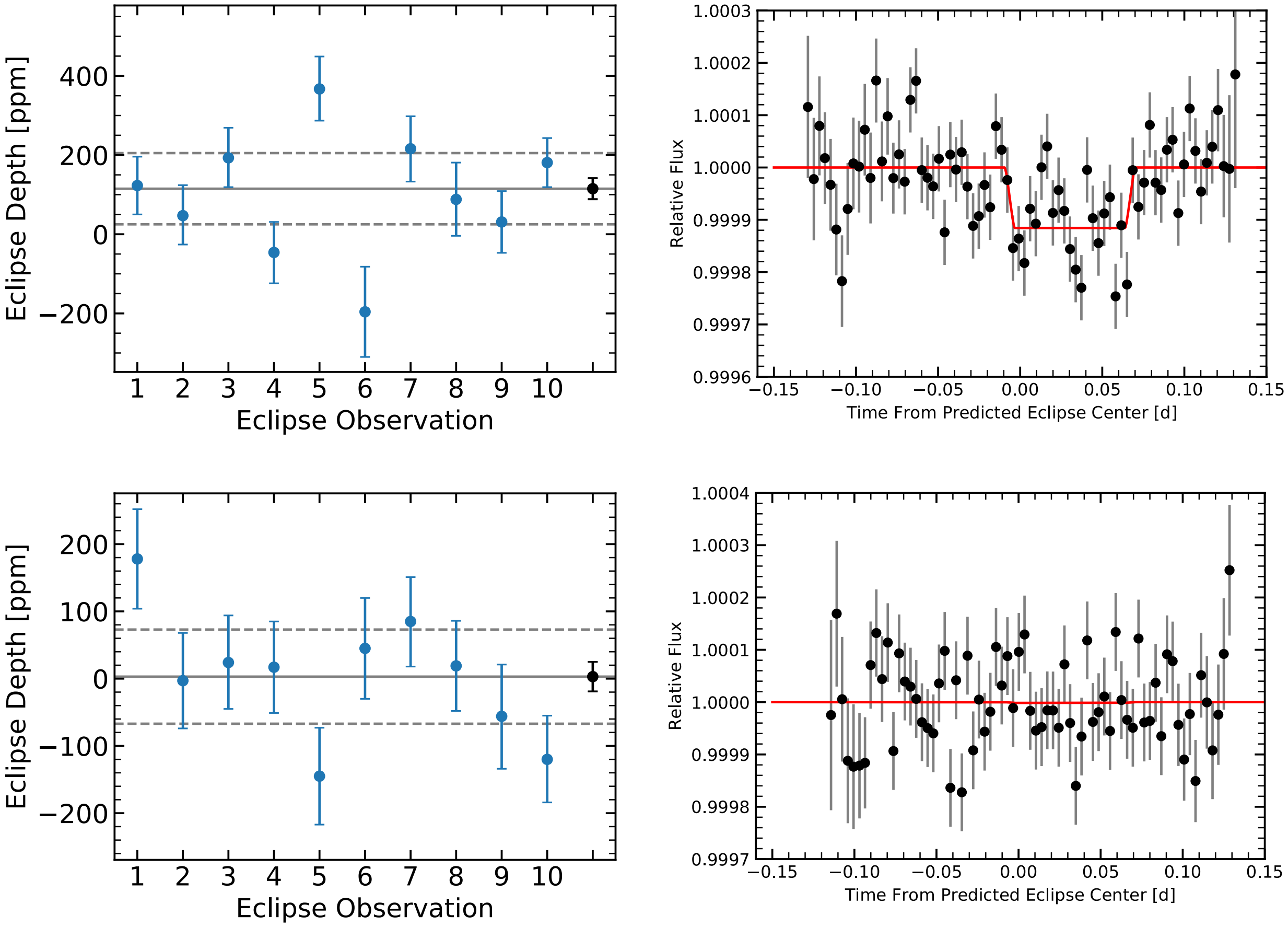

Supplementary Figure 5: Spitzer/IRAC secondary eclipse observations of GJ 3470b at 3.6 μm (top) and 4.5 μm (bottom). The left panels show the estimates of the eclipse depths for each of the ten individual eclipse observations (blue) and the global fit (black). Typical ±1σ uncertainties for individual eclipse fits are shown by gray dashed horizontal lines around the global fit value. The ten individual measurements are randomly distributed around the global fit. Consistent with the uncertainties, 6 of 10 and 7 of 10 data points are within the 68% confidence interval for the observations at 3.6 and 4.5 μm, respectively. The right panels shows the best fitting model light curves (red curve) from the global fit, overlaid with the systematics-corrected Spitzer data from all ten eclipse observations (black). The eclipse is slightly offset relative to the estimated eclipse time for a perfectly circular orbit. This is consistent with fits to the radial velocity data of GJ 3470b, which independently confirm a small but non-zero eccentricity (Kosiarek et al. 2018).

| Instrument/Grism | Wavelength [μm] | | | Depth [ppm] | +1σ [ppm] | -1σ [ppm] |
|---|---|---|---|---|---|---|
| **Transit:** | | | | | | |
| HST STIS G750L | 0.528 | – | 0.577 | 5912 | 178 | 192 |
| HST STIS G750L | 0.577 | – | 0.626 | 6135 | 104 | 104 |
| HST STIS G750L | 0.626 | – | 0.674 | 5999 | 89 | 94 |
| HST STIS G750L | 0.674 | – | 0.723 | 5945 | 91 | 91 |
| HST STIS G750L | 0.723 | – | 0.772 | 5937 | 98 | 107 |
| HST STIS G750L | 0.772 | – | 0.821 | 6215 | 85 | 86 |
| HST STIS G750L | 0.821 | – | 0.870 | 6093 | 108 | 110 |
| HST STIS G750L | 0.870 | – | 0.919 | 6098 | 113 | 119 |
| HST WFC3 G141 | 1.120 | – | 1.150 | 6101 | 34 | 35 |
| HST WFC3 G141 | 1.150 | – | 1.180 | 6050 | 37 | 34 |
| HST WFC3 G141 | 1.180 | – | 1.210 | 6077 | 33 | 34 |
| HST WFC3 G141 | 1.210 | – | 1.240 | 6035 | 31 | 32 |
| HST WFC3 G141 | 1.240 | – | 1.270 | 6126 | 34 | 31 |
| HST WFC3 G141 | 1.270 | – | 1.300 | 6082 | 31 | 29 |
| HST WFC3 G141 | 1.300 | – | 1.330 | 6055 | 29 | 29 |
| HST WFC3 G141 | 1.330 | – | 1.360 | 6196 | 26 | 28 |
| HST WFC3 G141 | 1.360 | – | 1.390 | 6178 | 31 | 30 |
| HST WFC3 G141 | 1.390 | – | 1.420 | 6111 | 32 | 32 |
| HST WFC3 G141 | 1.420 | – | 1.450 | 6127 | 29 | 31 |
| HST WFC3 G141 | 1.450 | – | 1.480 | 6136 | 31 | 32 |
| HST WFC3 G141 | 1.480 | – | 1.510 | 6124 | 32 | 31 |
| HST WFC3 G141 | 1.510 | – | 1.540 | 6079 | 26 | 29 |
| HST WFC3 G141 | 1.540 | – | 1.570 | 6121 | 31 | 32 |
| HST WFC3 G141 | 1.570 | – | 1.600 | 6057 | 34 | 30 |
| HST WFC3 G141 | 1.600 | – | 1.630 | 6059 | 33 | 35 |
| HST WFC3 G141 | 1.630 | – | 1.660 | 6061 | 32 | 32 |
| Spitzer/IRAC 3.6μm | 3.15 | – | 3.90 | 5763 | 64 | 65 |
| Spitzer/IRAC 4.5μm | 4.00 | – | 5.00 | 5941 | 60 | 66 |
| **Eclipse:** | | | | | | |
| Spitzer/IRAC 3.6μm | 3.15 | – | 3.90 | 115 | 27 | 26 |
| Spitzer/IRAC 4.5μm | 4.00 | – | 5.00 | 3 | 22 | 22 |

Supplementary Table 3: Transmission spectrum and eclipse depths

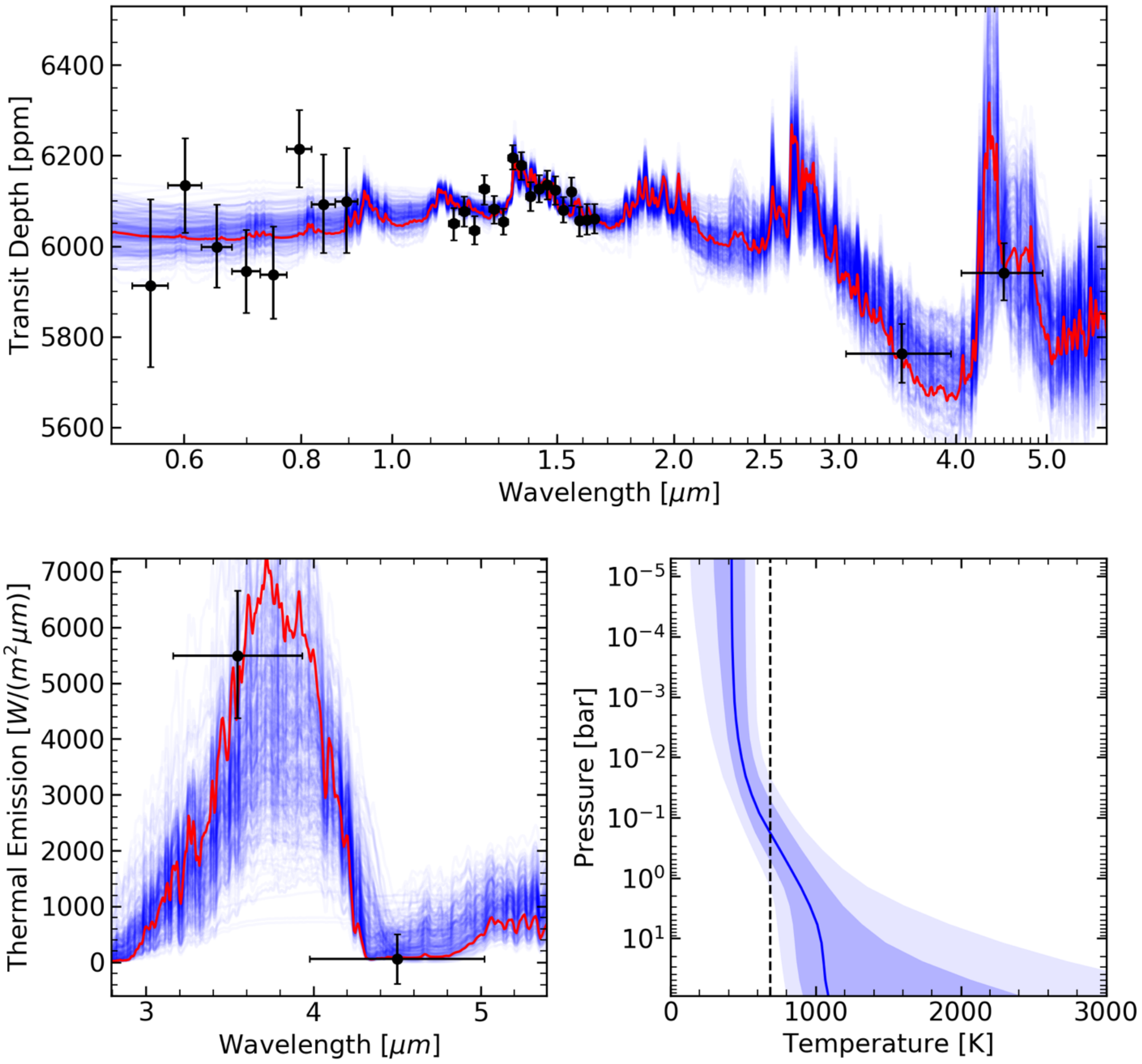

Supplementary Figure 6: Spectral fits and temperature profile constraints from the joint retrieval analysis of GJ 3470b's transit and eclipse data. The top and bottom left panels show the range of models fitting the observations (black) by depicting a random sample of 300 atmospheric models from the posterior distribution (thin blue curves). The best fitting model is shown as thick red curve. Consistent with the observed transit depth uncertainties, the models are tightly constrained within the precise *HST/WFC3* observations. At shorter and longer wavelengths, the *HST/STIS* and *Spitzer* observations allow for a slightly larger range of transit depths, mostly by slight changes in the cloud parameters as well as the CO and $CO_2$ abundances. The bottom right panel depicts the posterior constraints on the vertical temperature profile. The dark and light-blue shaded regions are the 1σ and 2σ spread in the temperature profiles, with the solid blue curve being the median temperature profile of all models in the posterior distribution. The equilibrium temperature for an planetary albedo of 0.1 is shown for comparison (dashed line).

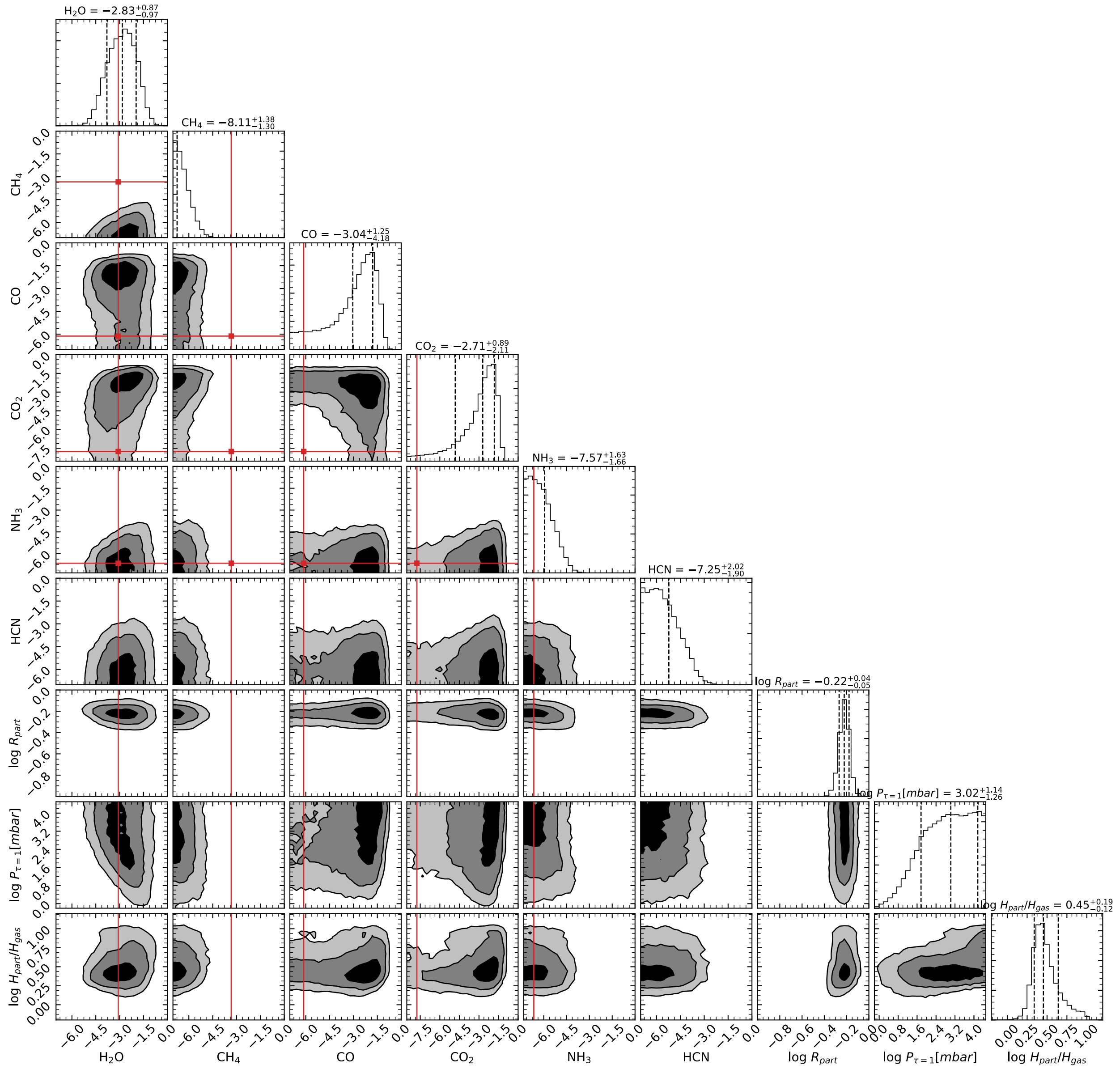

Supplementary Figure 7: Molecular abundance and cloud property constraints from the joint retrieval analysis of the transit and eclipse data. The top panels in each column show the 1D marginalized posterior distributions of the molecular abundances and cloud properties, with dashed vertical lines in the histograms indicating the marginalized 16th, 50th, and 84th percentiles. The subjacent 2D panels show the correlations among the gases and cloud properties, with the black, dark-gray, and light-gray regions corresponding to the 1σ (39.3%), 2σ (86.5%), and 3σ (98.9%) credible intervals. The vertical and horizontal red lines in each panel are the solar composition molecular abundances at 700 K and 0.1 bars, a representative photospheric temperature and pressure. The water mixing ratio is constrained to ±1 order of magnitude around 1 times solar. $CH_4$ and $NH_3$ are depleted. Note also the "elbow"-shaped correlation between CO and $CO_2$. This degeneracy arises because CO and $CO_2$ both absorb within the 4.5 μm Spitzer bandpass observed in transit and eclipse. Note that the retrieval included an additional 7 parameters for the vertical temperature structure and common-mode transit depth uncertainties which are not displayed here for clarity.

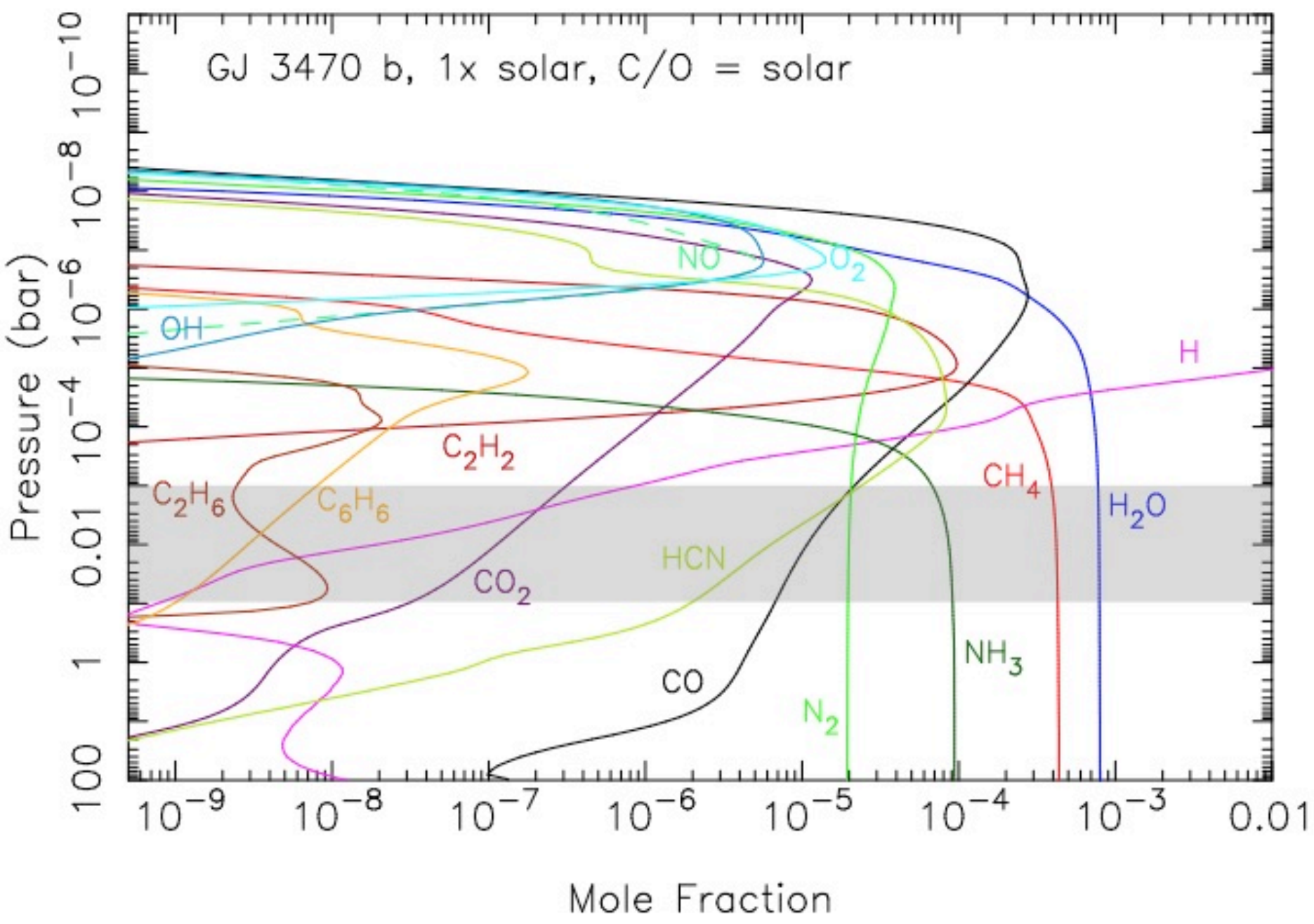

Supplementary Figure 8: Mixing-ratio profiles for several species of interest (as labeled) in our kinetics/transport models for GJ 3470b, for solar atmospheric metallicity and a C/O ratio of 0.54. The gray horizontal zone indicates pressure range to which our observations are most sensitive. The thermal emission observations extend to slightly deeper levels as well. Water and methane abundance follow mostly the equilibrium abundances, with photodissociation becoming relevant above approximately $10^{-5}$ bar, as previously also shown in a theoretical modeling investigation of the sub-Neptune GJ 1214b[15] and GJ 436b[16,21]. Ammonia is expected to be abundant the photosphere due to quenching resulting from vertical transport and the slow rate of reaction for the conversion to $N_2$ and $H_2$.

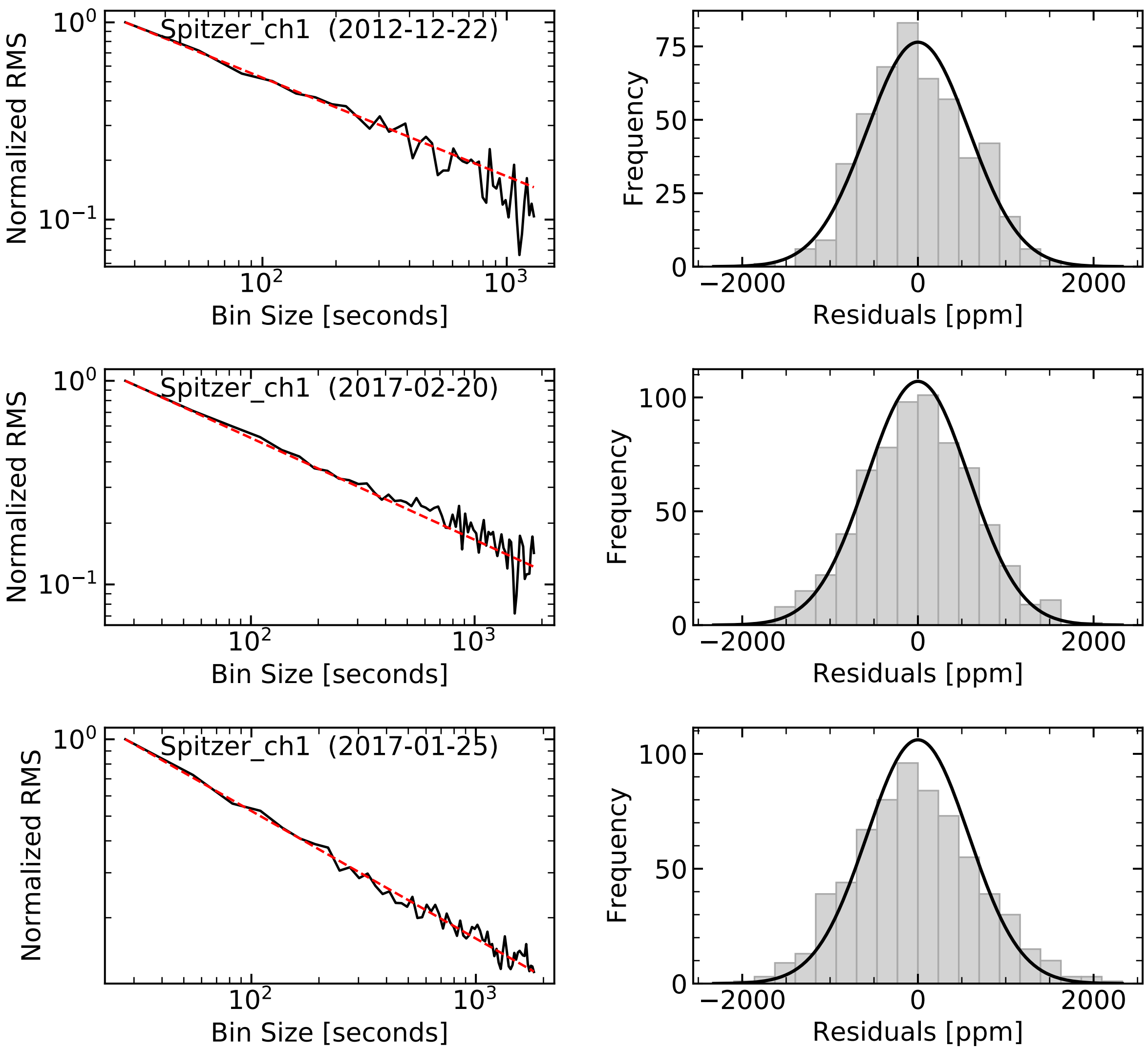

Supplementary Figure 9: Analysis of residuals from fitting the three Spitzer/IRAC 3.6um transits (top, center, bottom). Left panels: Photometric scatter vs. the width of the binning interval for Spitzer data. The root-mean-square error of the systematics-corrected Spitzer data (black) follows closely the theoretical square-root scaling for uncorrelated white noise (red dashed line), even when binned all the way to 30 minute intervals. Right panels: Histogram of the residuals (grey bars) compared to a theoretical Gaussian distribution with the width of the scatter parameter fitted as a nuisance parameter in the Bayesian analysis (black curve). The residuals are consistent with the Gaussian distribution and the scatter parameter.

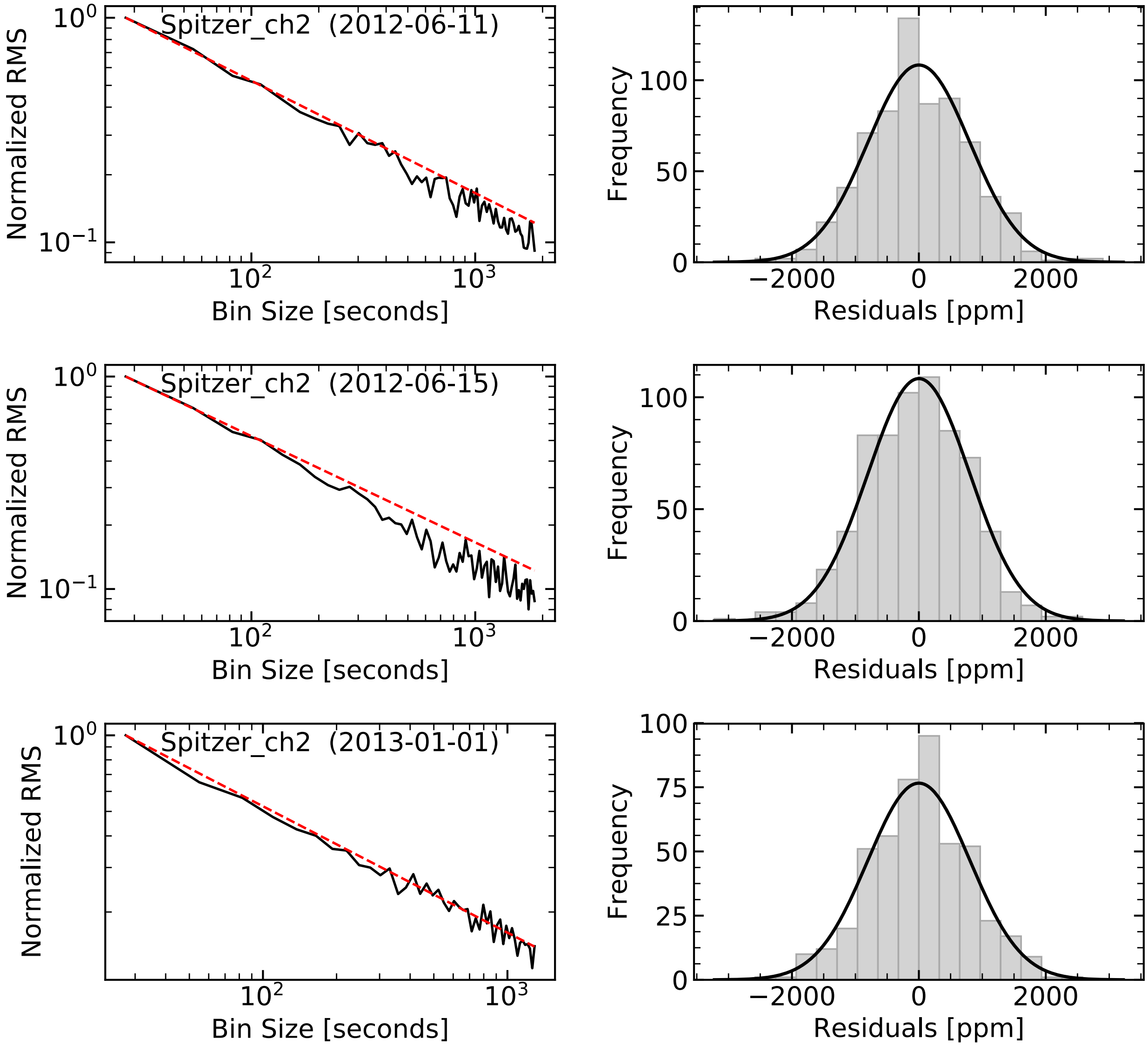

Supplementary Figure 10: Analysis of residuals from fitting the three Spitzer/IRAC 4.5um transits (top, center, bottom). Left panels: Photometric scatter vs. the width of the binning interval for Spitzer data. The root-mean-square error of the systematics-corrected Spitzer data (black) follows closely the theoretical square-root scaling for uncorrelated white noise (red dashed line), even when binned all the way to 30 minute intervals. Right panels: Histogram of the residuals (grey bars) compared to a theoretical Gaussian distribution with the width of the scatter parameter fitted as a nuisance parameter in the Bayesian analysis (black curve). The residuals are consistent with the Gaussian distribution and the scatter parameter.

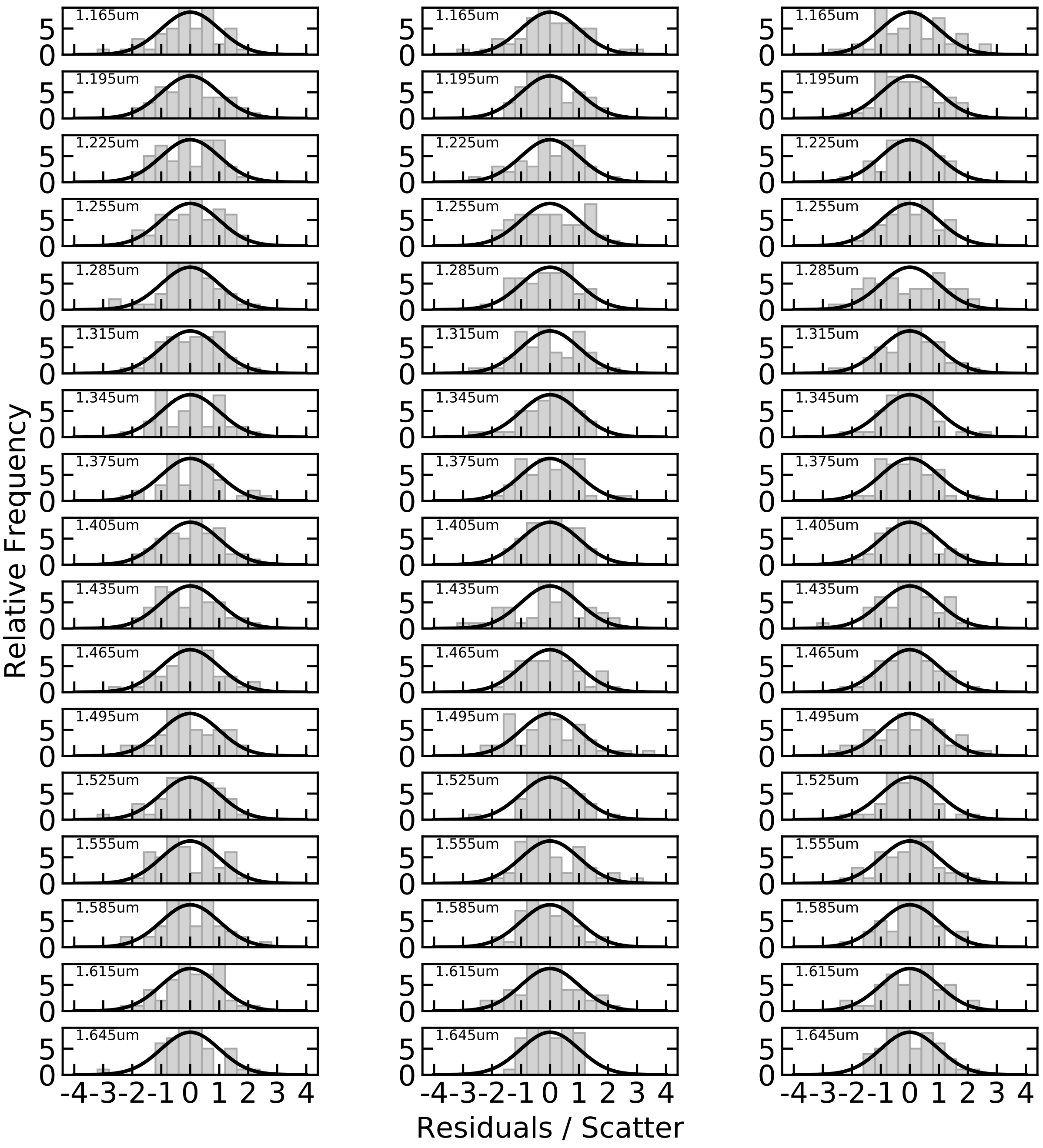

Supplementary Figure 11: Histograms of residuals from HST/WFC3 spectral light curve fits. This figure shows the histograms for each wavelength bin (rows) and each transit (columns) individually. Residuals are normalized by the fitted scatter parameter for the respectively wavelength bin and transit. The histogram of normalized residuals (gray bars) is compared to the normal distribution (black curve). Each histogram is made of only 51 data points leading to relatively poor sampling of each frequency distribution; however, no statistically significant deviation from the expected frequency distribution is observed. The agreement with the expected Gaussian distribution can be seen even better in the combined plot of all residuals shown in Supplementary Figure 12. WFC3 residuals are highly consistent with the Gaussian distribution and the fitted scatter parameter for each light curve.

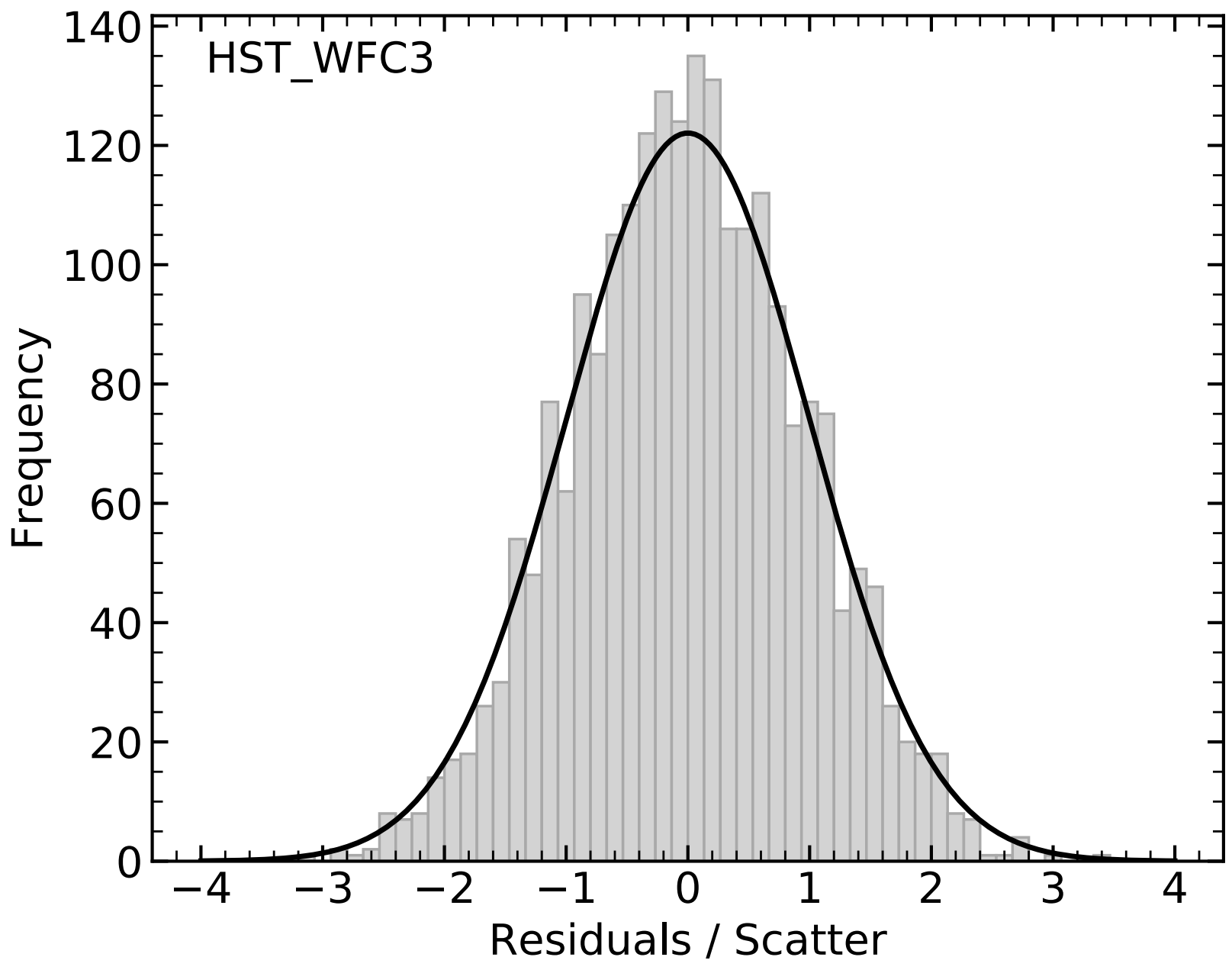

Supplementary Figure 12. Histograms of all residuals from WFC3 spectral light curve fits. This plot combines the residuals from all panels in Supplementary Figure 11 in order to increase the number of samples in the histogram. WFC3 residuals are highly consistent with the Gaussian distribution and the fitted scatter parameter for each light curve.

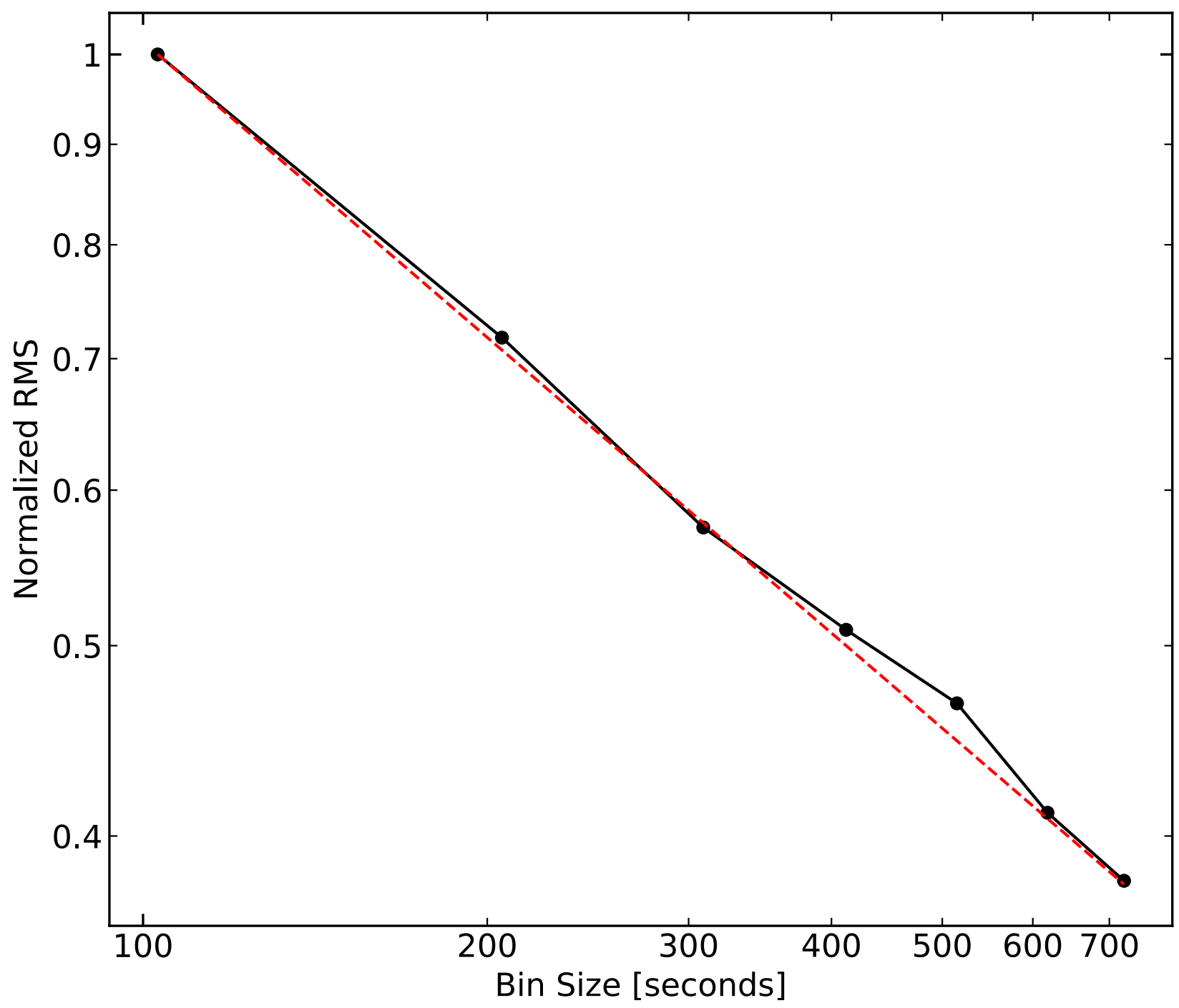

Supplementary Figure 13: Photometric scatter vs. the width of the time binning interval for all WFC3 spectroscopic light curves combined. The root-mean-square error of the systematics-corrected WFC3 data (black) follows closely the theoretical square-root scaling for uncorrelated white noise (red dashed line), even when binned all the way to 12 minute intervals. At this point only four data points are left per orbit. We conclude that time correlated noise is negligible.

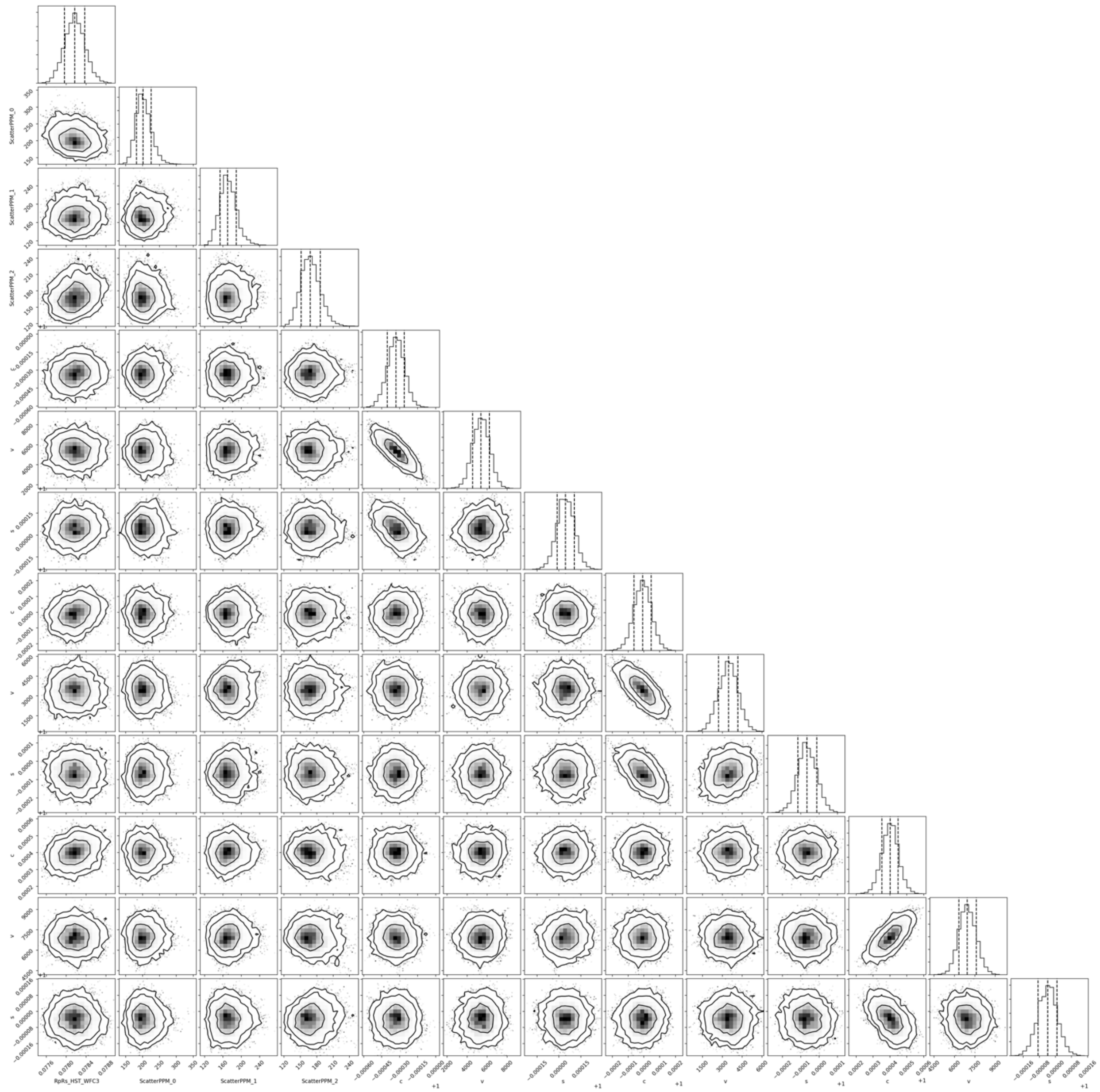

Supplementary Figure 14: Pairs plot showing the posterior distribution of the MCMC fitting parameters for the WFC3 spectral light curve fit. The panels on the diagonal show the marginalized posterior distribution for each fitting parameter. The 68% credible interval is marked by vertical dashed lines and quantified above the panel. The off-diagonal panels show the two-dimensional marginalized distribution for pairs of parameters, with the gray shading corresponding to the probability density and black contours indicating the 68% and 95% credible regions. Our instrument modeling results in no significant correlation between astrophysical transit depth (first column) and instrumental detrending parameters.

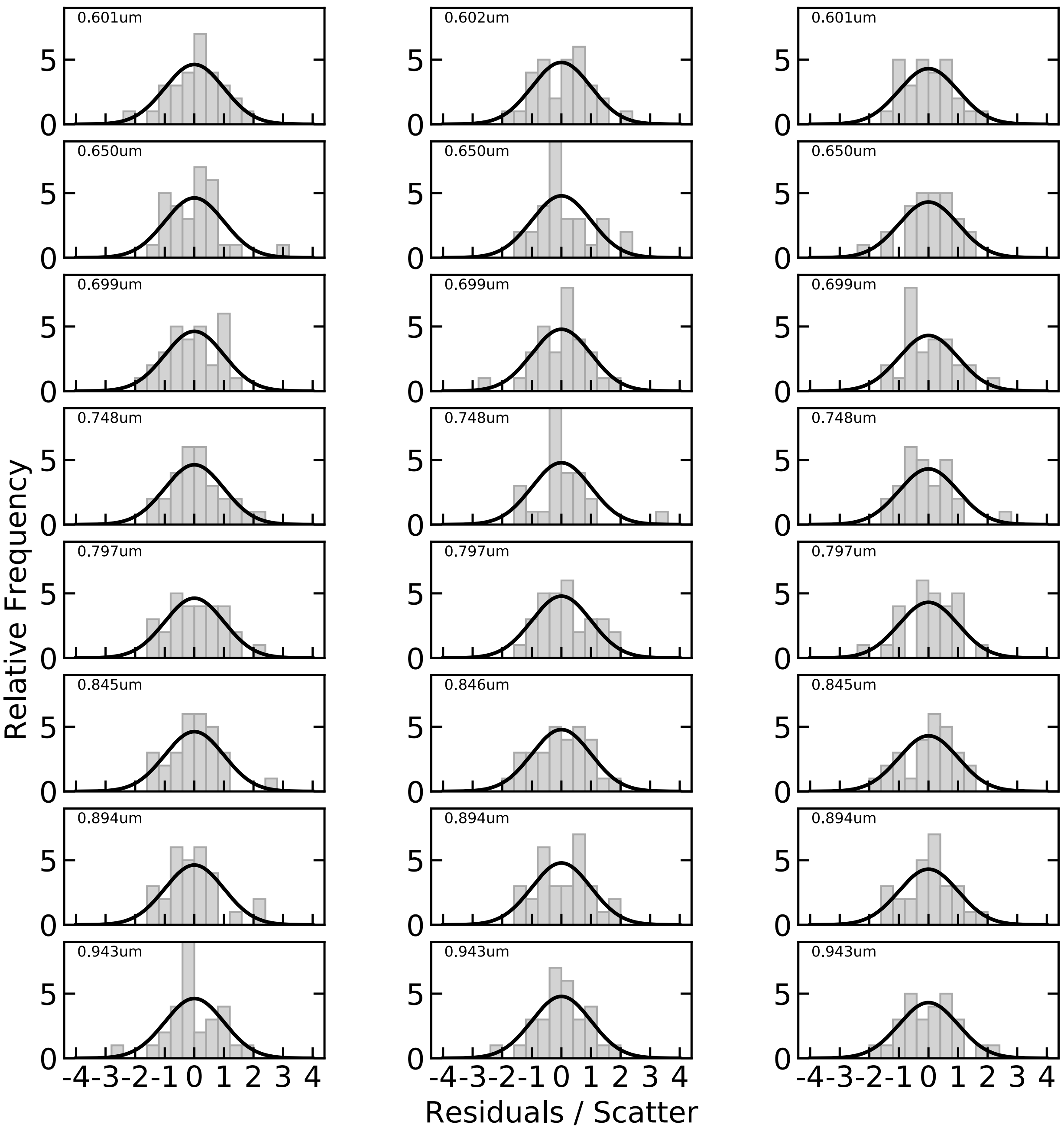

Supplementary Figure 15: Histograms of residuals from HST/STIS spectral light curve fits. This figure shows the histograms for each wavelength bin (rows) and each transit (columns) individually. Residuals are normalized by the maximum likelihood value of the scatter parameter for the respectively wavelength bin and transit. The histograms of the normalized residuals (gray bars) are compared to the normal distribution (black curve). Each histogram is made up of 29 data points leading to relatively poor sampling of each frequency distribution. All residuals combined are shown in Supplementary Figure 16. STIS residuals are consistent with a Gaussian distribution and the fitted scatter parameter is a conservative estimate of the scatter.

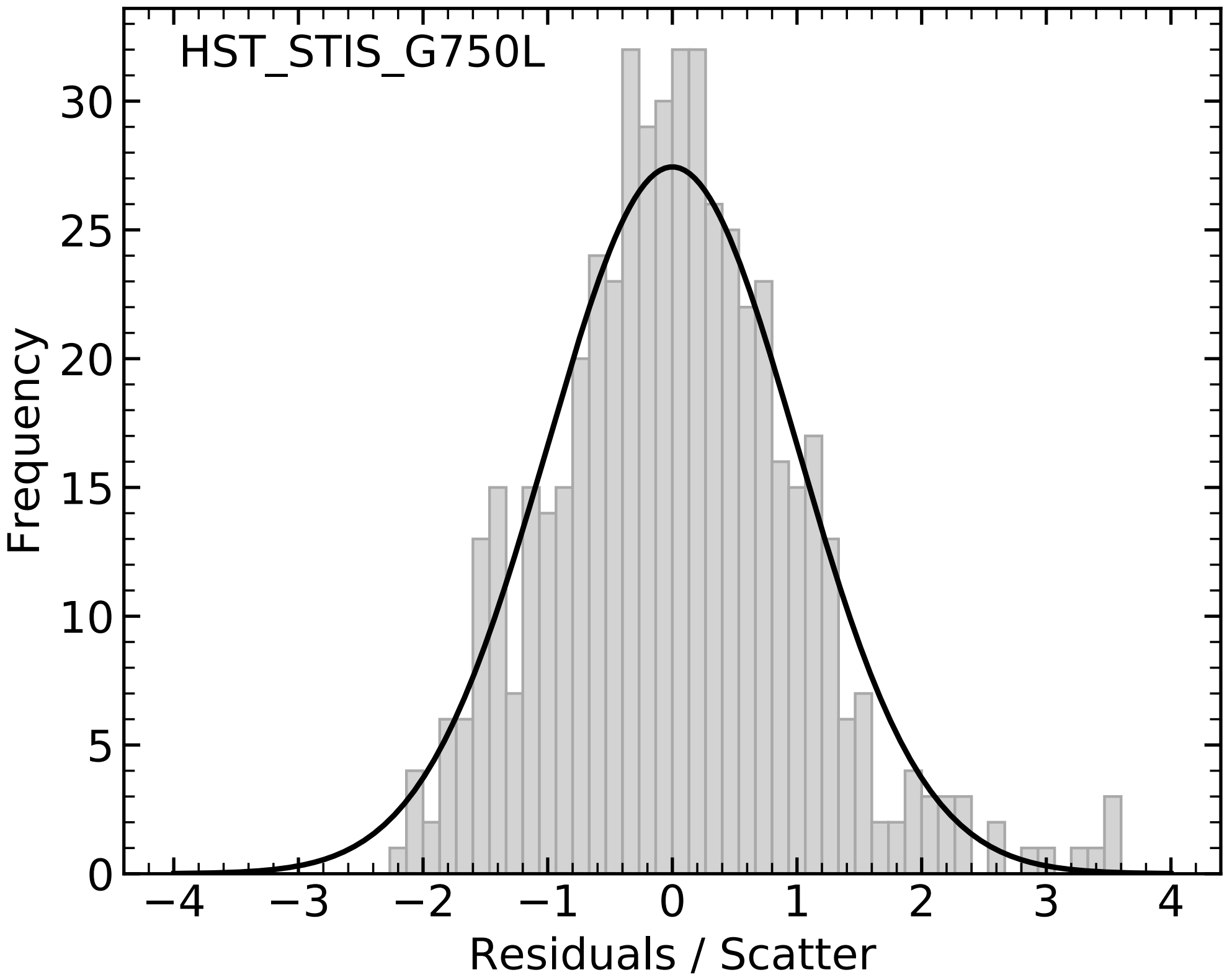

Supplementary Figure 16: Histograms of all residuals from STIS spectral light curve fits. This plot combines the residuals from all panels in Supplementary Figure 15 in order to improve the number of samples in the histogram. A distribution marginally narrower than the median of the scatter parameters is found. We conclude that our Bayesian analysis of the STIS light curves conservatively estimated the error bar as a result of the many detrending parameters needed to fit STIS light curves. This results in a conservative estimate of the transit depth uncertainties. Note that a standard maximum likelihood method would have found a smaller scatter because it would have estimated the scatter only based on the best fitting (potentially overfitting) model.

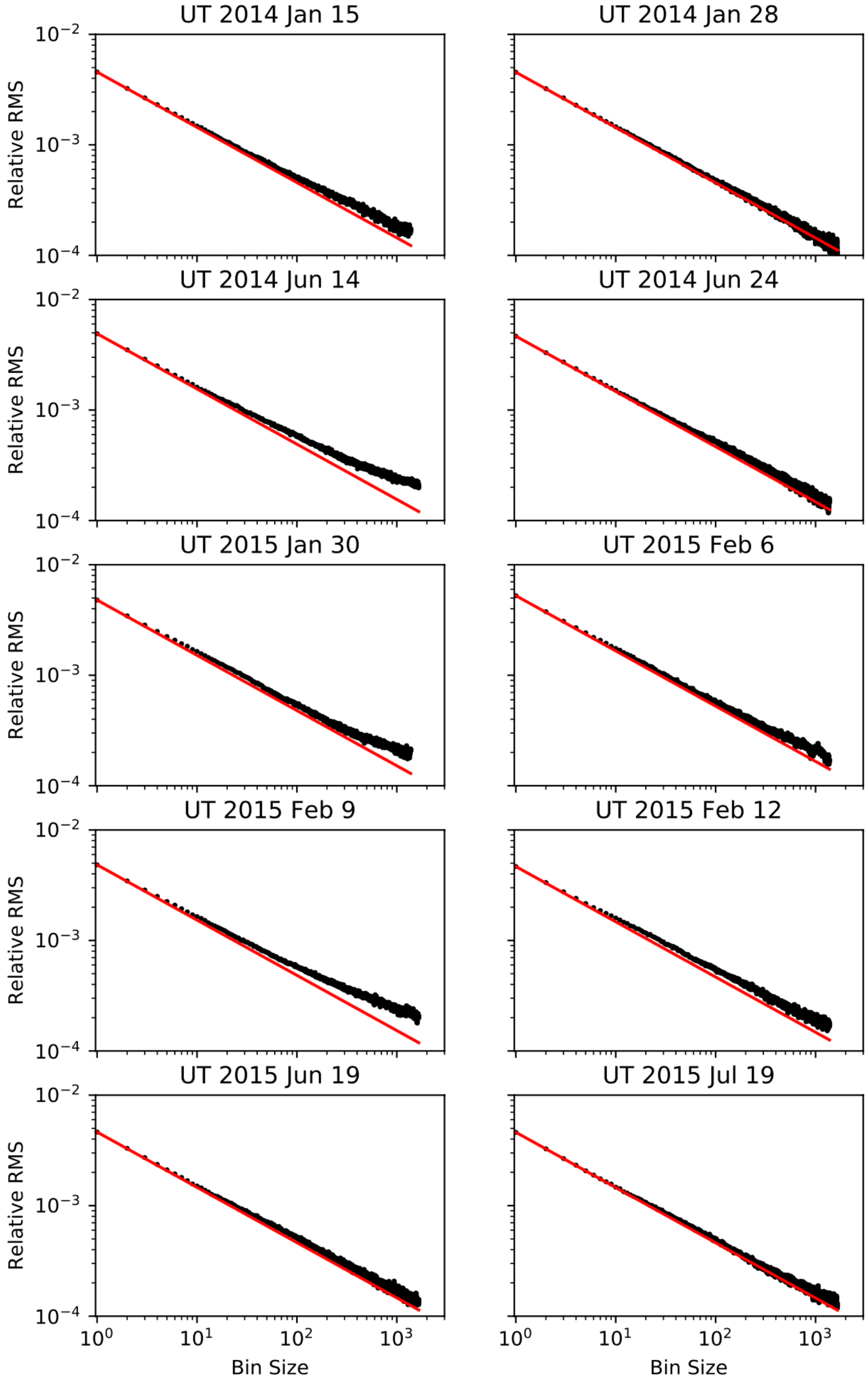

Supplementary Figure 16: Photometric scatter vs. the width of the binning interval for the ten Spitzer/IRAC 3.6um eclipses. The root-mean-square error of the systematics-corrected Spitzer data (black) follows closely the theoretical square-root scaling for uncorrelated white noise (red dashed line), even when combining up to 100 to 1000 data points to one bin.

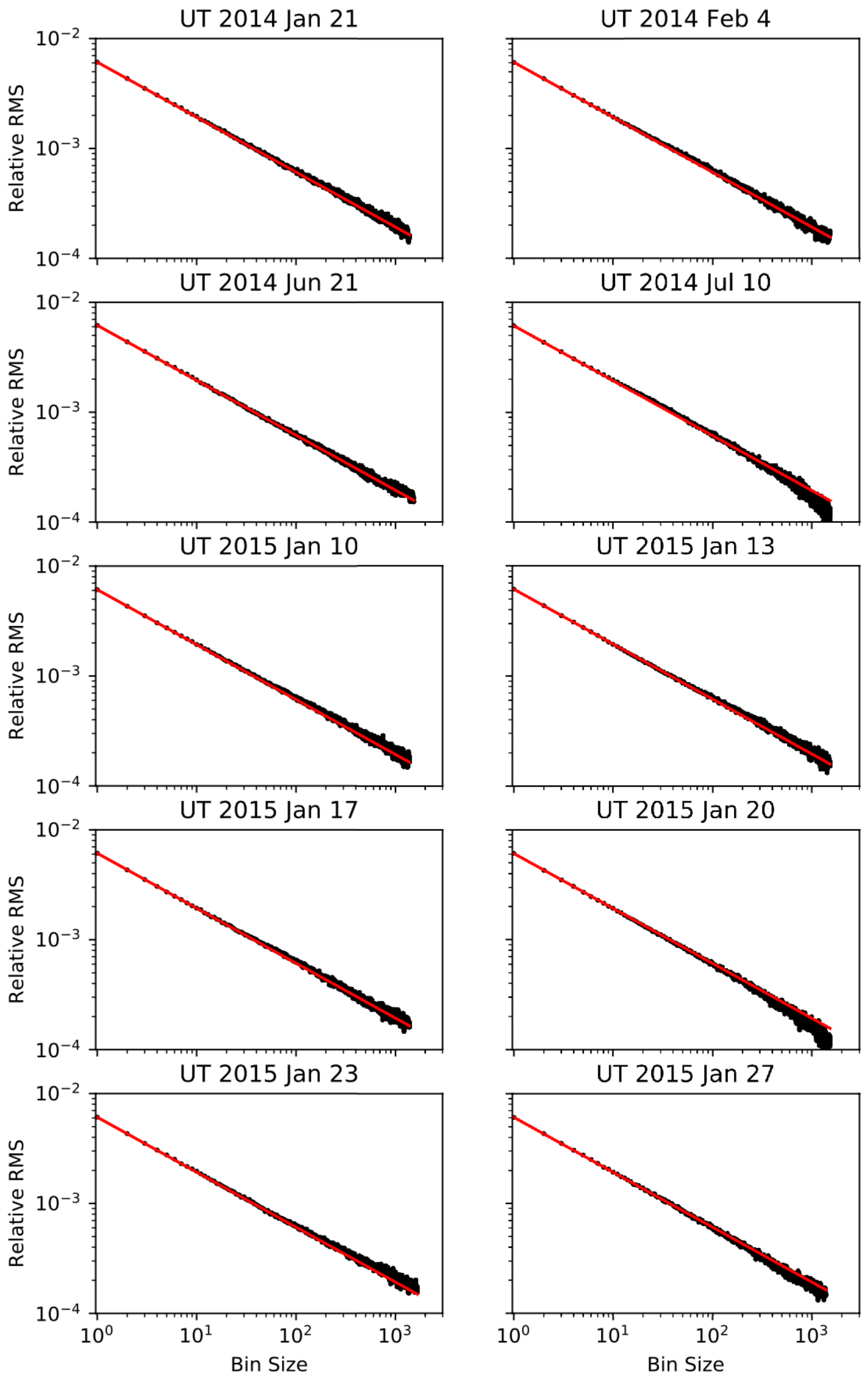

Supplementary Figure 17: Photometric scatter vs. the width of the binning interval for the ten Spitzer/IRAC 3.6μm eclipses. The root-mean-square error of the systematics-corrected Spitzer data (black) follows closely the theoretical square-root scaling for uncorrelated white noise (red dashed line), even when combining up to 100 to 1000 data points to one bin.